\documentclass{article}

\usepackage{arxiv}

\usepackage[utf8]{inputenc} % allow utf-8 input
\usepackage[T1]{fontenc}    % use 8-bit T1 fonts
\usepackage{hyperref}       % hyperlinks
\usepackage{url}            % simple URL typesetting
\usepackage{booktabs}       % professional-quality tables
\usepackage{amsfonts}       % blackboard math symbols
\usepackage{nicefrac}       % compact symbols for 1/2, etc.
\usepackage{microtype}      % microtypography
\usepackage{lipsum}

\usepackage{tikz}
\usepackage{amsmath}
\usepackage{amssymb,amsmath,epsfig,amsthm}
\usepackage{graphicx,latexsym,subfigure}
\usepackage{color,soul}
\usepackage{cite}
\usepackage{soul}
\usepackage{diagbox}

\usepackage{xifthen}
\usepackage{versions}
\usepackage{caption}

\newcommand{\heading}[1]{{\noindent\bf{#1}}~}
\newcommand\eat[1]{}

\title{DeviceWatch: Identifying Compromised Mobile Devices through Network Traffic Analysis and Graph Inference}

\author{
  Euijin Choo \\
  Qatar Computing Research Institute\\
  \texttt{echoo@hbku.edu.qa} \\
   \And
     Mohamed Nabeel  \\
  Qatar Computing Research Institute\\
  \texttt{mnabeel@hbku.edu.qa} \\
     \And
     Mashael Alsabah \\
  Qatar Computing Research Institute\\
  \texttt{msalsabah@hbku.edu.qa} \\
     \And
     Issa Khalil \\
  Qatar Computing Research Institute\\
  \texttt{ikhalil@hbku.edu.qa} \\
     \And
     Ting Yu \\
  Qatar Computing Research Institute\\
  \texttt{tyu@hbku.edu.qa} \\
     \And
     Wei Wang \\
  Beijing Jiaotong University\\
  \texttt{wangwei1@bjtu.edu.cn} \\
}

\begin{document}
\maketitle

\begin{abstract}
In this paper, we propose to identify compromised mobile devices from a network administrator's point of view. Intuitively, inadvertent users (and thus their devices) who download apps through untrustworthy markets are often allured to install malicious apps through in-app advertisement or phishing. We thus hypothesize that devices sharing  a  similar  set  of  apps will  have  a  similar  probability  of  being  compromised, resulting in the association between a device being compromised and apps in the device. Our goal is to leverage such associations to identify unknown compromised devices (i.e., devices possibly having yet currently not having known malicious apps) using the guilt-by-association principle. Admittedly, such associations could be quite weak as it is often hard, if not impossible, for an app to automatically download and install other apps without explicit initiation from a user. We describe how we can magnify such weak associations between devices and apps by carefully choosing parameters when applying graph-based inferences. We empirically show the effectiveness of our approach with  a comprehensive study on the mobile network traffic provided by a major mobile service provider. Concretely, we achieve nearly 98\% accuracy in terms of AUC (area under the ROC curve).
Given the relatively weak nature of association, we further conduct in-depth analysis of the different behavior of a graph-inference approach, by comparing it to active DNS data. Moreover, we validate our results by showing that detected compromised devices indeed present undesirable behavior in terms of their privacy leakage and network infrastructure accessed.
\end{abstract}

\section{Introduction}\label{sec_intro}

With the significant increase of online threats, there is a growing demand on ISPs by governments and organizations~\cite{australia-ISPs, australia2-ISPs} to have a bigger role in preventative cyber security. ISPs actively employ measures to filter spoofed traffic, but can also have a key role of detecting other attacks~\cite{wired-ISPs}. One emerging attack vector that can be effectively tackled at the ISP level is the detection of compromised mobile devices. Recently, researchers proposed to identify compromised devices from a ISP's point of view~\cite{social,carrier,cellbot}. ISPs have direct access to key network traces and information, which enables them to perform early detection of compromised mobile devices. Once discovered, ISPs can inform their customers including organizations so that they can take proper actions~\cite{darkside}.

Indeed, organizations have encouraged the use of personal mobile devices in workplaces, increasing the security incidents involving mobile devices. A recent study shows that one in three organizations has faced a security incident due to compromised devices having malicious apps~\cite{mobilesecindex2019}. Among other undesirable behavior, such devices may leak sensitive information, perform unauthorized credit card transactions, and phone calls~\cite{mosaic,taintdroid,recon,bugfix,cellbot}. A key challenge in mitigating such security threats is to accurately detect compromised devices and take actions. As organizations have little control over mobile devices and do not have access to all mobile network traffic, one needs to perform the detection at the mobile network provider level.

A number of methods to detect malicious apps have been proposed in the literature, which mainly apply various static and dynamic code analysis techniques~\cite{carrier,robotdroid,taintdroid,wild} and network-based approaches ~\cite{httpmalware,deviation,trafficav,machinelearning,imbalanced}. However, these techniques require the inspection of a vast number of apps created constantly and identify local features of every device and/or app. Therefore, a different method, that is not only robust but also scales to a large network, is required to detect compromised devices.

Many free apps are typically developed with in-app advertisements promoting other apps or in-app purchases ~\cite{appscanner}. While using such free apps, users are often tricked to authorize to download related apps and fall victim to \emph{drive-by-downloads} attacks~\cite{drive,phishing}. Further, many users also tend to install free apps that are not published in official app stores such as Google Play. For example, in some countries such as China, users are blocked from accessing Google Play and thus have to use various other stores with considerably low and varying security guarantees~\cite{beyondgoogleplay}. 
%We quickly discovered that people who install one abusive app are likely install many others as well ~\cite{symantec_creepware}. 

%devices with common apps \textcolor{red}{Issa: Devices with common apps??? I do not think this correct. Need to be rephrased} \nabeel{Nabeel: How about we change it to: so that unknown devices closely associated with many compromised devices are highly likely to be compromised.} will have a similar probability of being compromised. 
%alley: I changed it back to the original sentence
Motivated by the above observation, we hypothesize that there exists a \emph{homophily relationship} between devices and their installed apps so that devices sharing a similar set of apps will have a similar probability of being compromised. We thus formulate the compromised device detection problem as a graph-inference based classification task. Most apps require network connection between devices and their host servers while being downloaded, installed, or executed. We model such communication involving devices and mobile apps as a bipartite graph where one side is the set of devices (i.e., users) and the other is the set of mobile apps. To \emph{infer} whether an unknown device is compromised, 
%we explored several inference algorithms including Label Propagation [XXXXXX], Node2vec [XXXX], and Belief Propagation (BP). As we see later, BP has the best performance for our problem domain. BP is a well-known algorithm that has been widely used to reliably approximate an entity's likelihood of being bad on probabilistic graphical models for large graphs in a variety of security contexts, including anomaly detection, fraud detection, and malicious domain detection ~\cite{maldomain,fraudeagle,generalbp}.
%\textcolor{blue}{Issa: I rewrote the below paragraph. Just add the references to replace the XXXXXXs above}
having evaluated several graph inference approaches, we apply Belief Propagation (BP), a well-known algorithm that has been widely used to reliably approximate an entity's likelihood of being bad on probabilistic graphical models for large graphs in a variety of security contexts, including anomaly detection, fraud detection, and malicious domain detection ~\cite{maldomain,fraudeagle,generalbp,enterprise_bp}.

%Anecdotal evidence suggests that those users who often download apps from non-official, risky app stores are more likely to download other malicious apps from such app stores.

%While an app may not necessarily have a direct relationship with another published app in the same store (e.g., repackaged apps or developed by the same developer), the maliciousness of an app can be related to others because of the maliciousness of the app store or shared malicious advertisement servers~\cite{unsafe,robust}. In this paper, we thus define \textbf{compromised devices} as ones whose owners often inadvertently download malicious apps and have high probability to download other malicious apps. The goal of this research is then to identify such compromised devices by leveraging the association between compromised devices and malicious apps installed in them.

%There are several fundamental questions that we investigate: (1) Is there any association between a device being compromised and apps installed in the device.
%(2) If any, would a graph-inference approach based on such association be effective to identify other compromised devices given the association? (3) Is there any difference/similarity between the association in mobile networks and in other types of networks such as domain-IP graphs ~\cite{maldomain}? If any, how would the difference/similarities affect the behavior of graph inference algorithms when applied to each application?

Essentially, the effectiveness of BP depends on the strength of association between nodes in the graph ~\cite{codaspy_issa}. Unlike other applications where associations are relatively straightforward to be derived (e.g., a malware-infected machine and its activity controlled by command \& control servers%malware writers
), it is quite challenging to derive such strong associations between devices and mobile apps due to the facts that: (1) it is often hard for mobile apps to interfere and taint other apps; (2) user interactions are needed to take any action. 

We empirically verify our hypothesis that, to a certain extent, there in fact exist associations which can be used to correctly identify compromised devices using 5-terabytes of anonymized mobile network dataset provided by a cellular service provider. We further discuss the effect of the relatively weak associations in a device-app graph on BP. Concretely, we provide in-depth analysis on the topological similarity and differences between a device-app graph and well-known domain-IP resolution graphs obtained from active DNS data~\cite{codaspy_issa} and their impact on the behavior of BP. Finally, we investigate if detected devices exhibit undesirable behavior in terms of privacy leakage and hosting servers accessed by the devices. 

In sum, we make the following main contributions. First, we investigate an association between mobile devices and apps installed in the devices with which network administrators can successfully identify \textbf{unknown compromised devices} with little knowledge on devices in the network. Then, we model the associations as a device-app bipartite graph and apply a graph-based inference approach using the association. To the best of our knowledge, this is the first attempt to use such associations to detect compromised devices in mobile context. The key advantage of our approach 
is that (1) it is applicable regardless of devices' models, OS versions, app 
versions, or app types (e.g., phishing, malware) and (2) it can detect compromised devices at large-scale without time-consuming investigation on individual devices. Second, through experiments over a large-scale real-world dataset, we show that our approach can effectively detect compromised devices, achieving nearly 98\% accuracy in terms of AUC. (3) We further investigate the unique graph structures of the association between devices and their installed apps, and how it affects the choice of key parameters of BP and the effectiveness of our approach. (4) Finally, we validate our approach with a post-analysis of the behavior of detected unknown compromised devices. We show that these devices, most of whose apps are not known malicious apps, are leaking highly sensitive information in their network traffic, and they tend to frequently access IPs and domains with malicious behavior such as fast-fluxing or being short-lived ~\cite{fastflux:2008:NDSS,takedown:2019:NDSS}. 

\section{The Proposed Approach}\label{sec_approach}

%The aim of this research is to identify additional vulnerable mobile devices given a set of known ones and the network data of mobile devices collected by a network administrator or a mobile service provider. To do so, there are several challenging problems to be addressed. In the following, we first discuss the challenges we encounter in this problem and provide the overview of our approach to tackle each challenge. 

The aim of this work is to identify unknown compromised mobile devices given a small set of known ones and the network traffic data of mobile devices collected by a service provider. Concretely, we want to determine whether a given unknown device is compromised or not by analyzing their connections to known devices. We first highlight the main questions and challenges that we must address and describe how our approach tackles each challenge.

%who are likely to download malicious apps based on the current set of apps installed in the devices. 

\subsection{The Baseline Approach}~\label{blacklist}

One trivial approach to detect compromised devices is to compare apps in a device against known malicious apps. However, similar to other blacklist based approaches utilized to detect malicious entities in the Internet, such an approach fails to detect compromised devices having previously unknown malicious apps ~\cite{impending}. One thus needs approaches that can predict the status of devices based on the limited prior knowledge. In this paper, we thus propose to employ a graph inference approach, presented in the following.

\subsection{The Graph-Inference Based Approach}\label{overview}
 %In this research, we investigate two types of associations: (1) a device-IP association, and (2) a device-app association. %\mashael{If we want to talk briefly about the other graph approaches to address point A1 that Nabeel is experimenting with, or refer to the extended version or discussion, here is a good place.} Nabeel: I think we don't need to discuss about other approaches here.

%The first challenge is that based on which association we can connect devices together and how to model such association.
A key challenge to identify compromised devices using an inference algorithm is to first identify meaningful associations that can graphically demonstrate the homophily relationship. That is, we seek to define an association that is able to create two distinct clusters in the association graph, corresponding to compromised and not-compromised devices, respectively.

Mobile users mostly access the contents through the apps in their devices~\cite{robust}. We thus investigate the associations between devices and apps. Intuitively, if a device installs malicious apps, it is likely to download/install other malicious apps due to several reasons including in-app advertisements promoting similar apps~\cite{appscanner} and other drive-by-download attacks. Meanwhile, most apps require network connection between devices and their host servers while being downloaded, installed, or executed. We thus claim the likelihood of a device being compromised can be measured by analyzing its app usage behavior revealed in the network traffic. Our key insight is thus that there exists an association between a device and apps following homophily that can be used to identify other compromised devices. 

We present a model to reflect the homophily relationship between a device and apps. Specifically, we capture the association between devices and their apps through analysis of network traffic (as an ISP has no direct access to devices) and model such associations as a bipartite graph (Section~\ref{sec_bipartite}).

To determine whether a device is compromised, we follow the guilt-by-association principle that has been extensively applied in various applications with a graph model~\cite{codaspy_issa,fraudeagle,guiltfile,impending,maldomain}. In a nutshell, the idea of guilt-by-association is to estimate the \emph{guiltiness} of a node by propagating the prior knowledge on some of the nodes in the graph model, given the homophily relationship between nodes.

One may utilize various approaches such as label propagation (LP)~\cite{lp:2002,marmite}, BP, or graph node embedding~\cite{node2vec} to perform inference over graphs. We show in Section~\ref{discussion} that the accuracy of these approaches is comparable with BP being slightly better than the other approaches and much more efficient than graph node embedding. Hence, we apply BP on the bipartite graphs we build (Section~\ref{sec_bp}).

\subsubsection{Constructing Bipartite Graphs}\label{sec_bipartite}
\begin{figure}[tb]
\begin{center}
\parbox{0.45\textwidth}{
\centering
    \epsfig{file=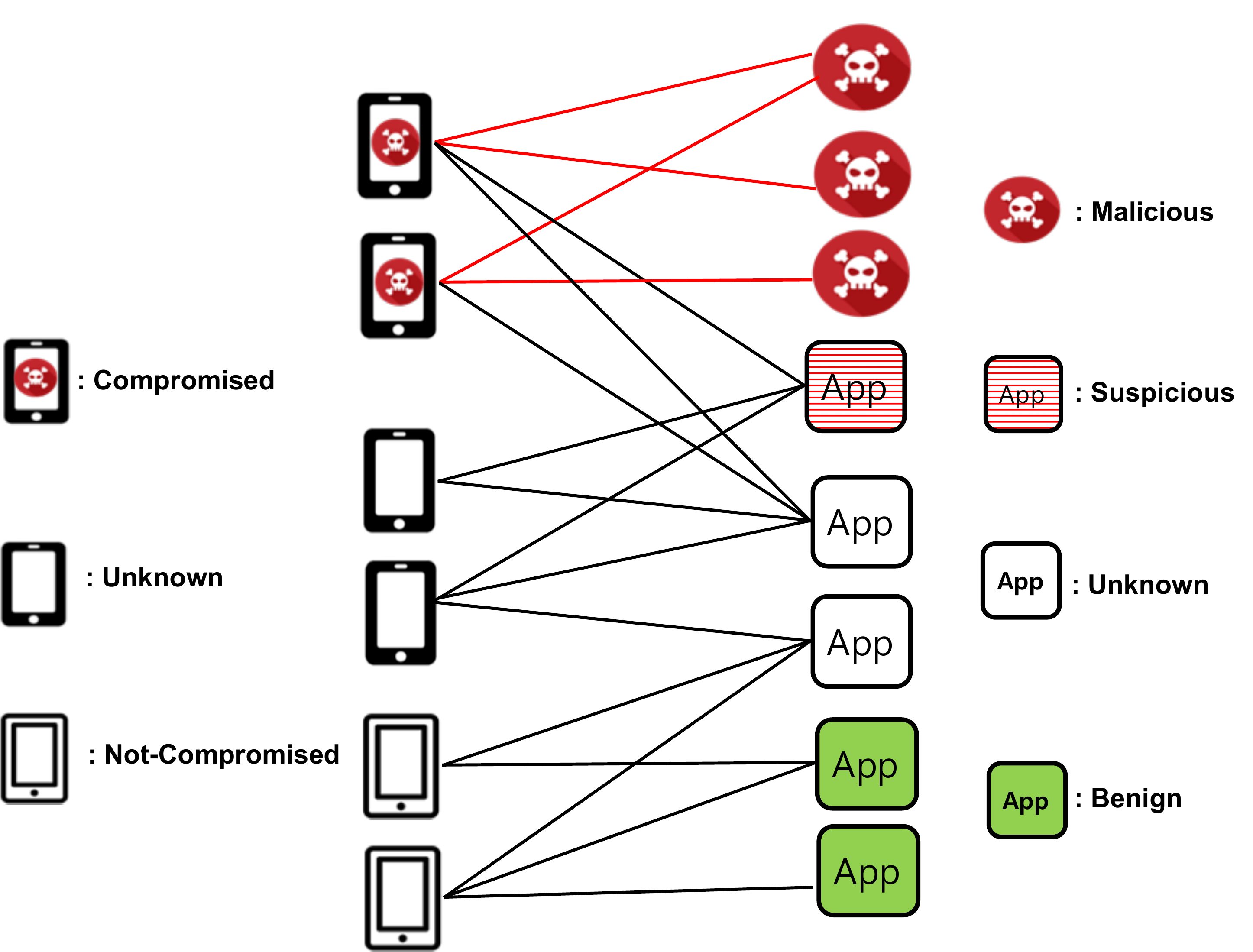,width=0.3\textwidth}
\caption{A device-app bipartite graph model}\label{bipartite}}

\end{center}
\end{figure}

We represent the associations between devices and apps as a bipartite graph $G = (V,E)$ where a set of devices $D = \lbrace d_1,...d_n\rbrace \subset V$ and a set of apps $A = \lbrace a_1,...a_n\rbrace  \subset V$ are connected with undirected edges $e(d_i,a_j)$, where $d_i$ is a device and $a_j$ is an app. Fig.~\ref{bipartite} illustrates an example of a bipartite graph model where the left side is a set of devices (i.e., nodes in $D$) and the right side is a set of apps (i.e., nodes in $A$). Each node in $D$ may belong to one of three categories: not-compromised, compromised, and unknown, and each node in $A$ may belong to one of four categories: benign, malicious, suspicious, and unknown. As illustrated, a compromised device may have edges with all types of apps including malicious, suspicious, unknown, and benign. An unknown device may have edges with suspicious, unknown, and benign apps. A not-compromised device may have edges with benign and unknown apps, but does not have edges with suspicious or malicious apps.

%Our approach follows the guilt-by-association  principle that has been extensively applied in various domains with a graph model~\cite{codaspy_issa,fraudeagle}. In a nutshell, guilt-by-association is a method to estimate the maliciousness of a node by propagating the prior knowledge on part of nodes in the graph model. 

%(3) Research problems: (1) admittedly the association between apps are not as strong as in other contexts (e.g., domains visited by infected hosts or files downloaded by infected machines), as it must involve a user to be tricked and to take some actions. It is a question whether such a weaker association is effective to infer vulnerable devices without being overwhelmed by other noises (e.g., popular apps that are installed in almost all devices); (2) compared to the spaces of domains (the HP paper) and downloaded files (the Georgia Tech paper), the domain of apps is much smaller. So the chance that two devices install same apps is much bigger, which means it is more likely that vulnerable devices are ``closer'' to non-vulnerable devices, making inferences harder to tell them apart. 

% alley: this is about intuition about association. Eventually, bp score is the probability. P(Bad) means the probability of being compromised
%\nabeel{Nabeel: is this intuition correct? When we collect ground truth we assume that compromised devices have malicious apps. Hence, there is no probability involved because of our assumption. Should we omit this sentence?}

\subsubsection{Belief Propagation}\label{sec_bp}

We work based on the intuition that there exists an association between the probability of a device being compromised and the probability of having malicious apps. That is, the more malicious apps a device has, the more likely it is to download other malicious apps, resulting in  homophily relationships between devices and apps. 
%\nabeel{Nabeel: Should we rephrase this sentence to match what we have mentioned about homophily relationships in abstract and intro? Something like: We work based on the intuition that devices currently having malicious apps are likely to download other malicious apps, resulting in  homophily relationships between devices and apps.}
Given the bipartite graph in Section~\ref{sec_bipartite} and the prior knowledge about devices, we thus aim to infer the probability of unknown devices being compromised or not using the guilt-by-association principle. To do so, we employ BP, which is shown to reliably estimate the posterior probabilities on probabilistic graphical models for large graphs in a variety of domains including anomaly detection, fraud detection, and malicious domain detection~\cite{guiltfile,codaspy_issa,fraudeagle}. In the following, we explain how we apply BP in our context.

We model each node  $i \in V$ as a random variable, $x_i$, that can be in the set of \emph{states} $S = \lbrace good , bad \rbrace$ so that the badness and goodness of a node can be expressed by the probabilities $P(Bad)$ and $P(Good)$, respectively, where $P(Bad) + P(Good)= 1$. Our goal is then to determine the marginal probabilities $P(x_i = Good)$ and $P(x_i = Bad)$ for unknown devices. 

BP computes the marginal probability of each node by iteratively passing local messages from its neighbor given the prior knowledge of other nodes in the graph.%, which is defined as follows:
%Nabeel: the marginal probability of a given node, say X_i, can be computed by taking the  sum  over  all  possible  states  of  all  the  other  nodes  in  the graph.  The  complexity  of  this  equation  is  exponential  in  the number  of  nodes. BP is an approximation to this which works in the worst case $O(V^2)$.
\begin{comment}
\begin{equation}\label{eq:pro}
\tag{Eq.1}
P(x_i) = \sum_{x_1}\cdots\sum_{x_{i-1}}\sum_{x_{i+1}}\cdots\sum_{x_n} P(x1,x2,\cdots,x_n)
\end{equation}
\end{comment}

At each iteration, BP computes the message vector  $m_{ij}$ for each node $i$, and passes it to each of its neighbors $j \in N(i)$, where $N(i)$ is the set of $i$'s neighbor. $m_{ij}(x_j)$ is $i$'s belief  that node $j$ is in state $x_j$ (i.e., $i$'s outgoing message vector to $j$), which will be computed based on $i$'s neighbors' messages about $i$. Concretely, there are three components to compute message $m_{ij}(x_j)$: (1) initial belief $\phi_i(x_i)$ for $i$ being in state $x_i$; (2) the product of all messages from $i$'s neighbors excluding $j$ (i.e., $i$'s incoming message vector from $k \in N(i)$); and (3) the edge potential $\psi_{ij}(x_i,x_j)$ between two neighboring nodes $i$ and $j$ specifying the probability of $i$ being in state $x_i$ and $j$ being in state $x_j$. Formally, the message $m_{ij}$ is defined as:
\begin{equation}\label{eq:msg}
\tag{Eq.1}
m_{ij} (x_{j}) = \sum_{x_i \in S} [\phi_i(x_i)\psi_{ij}(x_i,x_j) \prod_{k \in N(i)\backslash j}m_{ki}(x_i)]
\end{equation}

We assign the initial belief for each node based on the ground truth labels, which is summarized in Table~\ref{bp_table}(a). Furthermore, Table~\ref{bp_table}(b) represents the edge potential matrix. In Section~\ref{sec_analysis}, we will discuss how we choose $\delta$ and $\epsilon$, and the effect of varying these parameters on detection accuracy.

\begin{table}[h!]
\begin{center}
\subtable[Initial beliefs for nodes in a graph\label{initial}]{
 \begin{tabular}{|c|c|c|}\hline
\textbf{ }  & P(Bad)& P(Good)\\
\hline
Bad        & $\delta$ & 1- $\delta$\\
\hline
Good       & 1- $\delta$ & $\delta$\\
\hline
Unknown       & 0.5 & 0.5\\
\hline
\end{tabular}}
\subtable[Edge potentials\label{edge}]{
\begin{tabular}{|c|c|c|}\hline
\textbf{ }  & Bad& Good\\
\hline
Bad        &  $\epsilon$ & $1-\epsilon$\\
\hline
Good       & $1-\epsilon$ & $\epsilon$\\
\hline
\end{tabular} }
   \caption{Initial beliefs and edge potentials for Belief Propagation }
   \label{bp_table}
  \end{center}
\end{table}

%Nabeel: Commented out the following as we do not need to explain the this. Non-convergence for certain cases is a known issue. 
%alley: the few sentences in the beginning is BP stops, when it converges. Shouldn't we talk about the condition of stops? as we also talk about convergence later. what will it mean message converge  below?
%Nabeel: In the next paragraph, we say convergence is when messages do not change between subsequent iterations. isn't it enough? Also, I forgot to mention above that, aynchronous or synchronous messaging details about BP is also quite well known.
\begin{comment}
Following~\cite{maldomain}, we update $i$'s outgoing messages in \emph{a synchronous order} for simplicity. That is, the $i$'s outgoing messages in iteration $t$ is computed from $i$'s incoming messages in iteration $t-1$. And the message passing procedure stops when the messages converge, i.e, they do not change significantly between iterations.  Note that although BP is not theoretically guaranteed to converge given arbitrary graph topologies, it is shown to converge quickly with highly accurate approximation in practice~\cite{guiltfile,maldomain}.
\end{comment}

Note that BP is not theoretically guaranteed to converge for arbitrary graphs. However, it is shown to converge quickly with highly accurate approximation in practice~\cite{guiltfile,maldomain}. After the messages converge, i.e, they do not change significantly between iterations, we compute the final belief for $i$ as follows:

\begin{equation}\label{eq:belief}
\tag{Eq.2}
b_i(x_i) = C\phi(x_i) \prod_{k \in N(i)}m_{ki}(x_i),
\end{equation}

where $C$ is a normalizing constant. Finally, we classify devices as compromised or not based on the final belief.

\section{Dataset}\label{sec_data}

\subsection{Mobile Network Traffic Dataset}\label{sec_ISPdata}

Our dataset contains 5-terabytes of 5-days mobile network traffic data from a Chinese mobile service provider. Note that ISP does not have any control over the apps running on devices so that it is not straightforward to identify which devices use which apps and build a device-app bipartite graph from the traffic. In the following, we describe how we identify and extract information about devices and apps from the dataset.

\begin{figure}[tb]
\begin{center}
\parbox{0.4\textwidth}{
\centering
    \epsfig{file=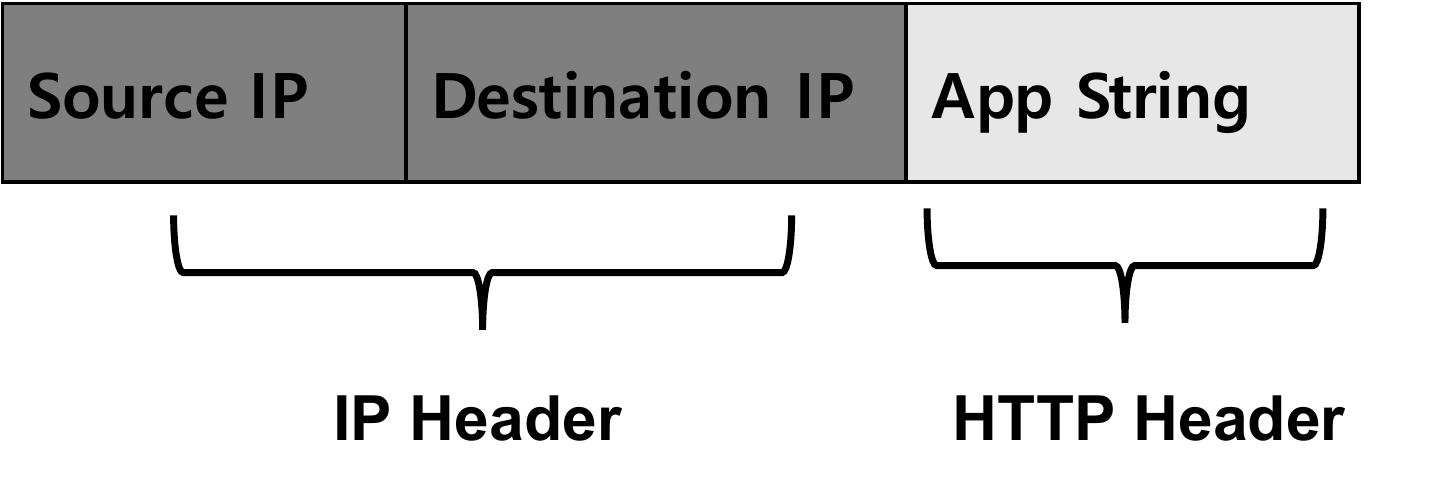,width=0.28\textwidth}%\vspace{-.21in}
\caption{Simplified IP packet with fields used for data extraction}\label{fig:fields}}
\end{center}

\end{figure}

Our approach relies on constructing bipartite graphs using entities extracted from the mobile network traffic including devices and apps. Fig.~\ref{fig:fields} shows the fields we use from IP packets to extract those entities. The \emph{app string} and destination IPs are  important in extracting app information; whereas the source IPs are important to  extract device information. 

\subsubsection{Device Extraction}\label{sec:device_extraction}

Following previous research~\cite{appprint,carrier,mosaic,appscanner}, we consider each source IP address as a device. By doing so, we extract 250149 devices from our traffic.

\subsubsection{App Extraction}\label{sec:app_extraction}

Given the ISP traffic, one may utilize various information to extract app information such as app strings in the HTTP header, destination IP addresses in IP header, or unique TCP traffic patterns representing apps~\cite{appprint,appscanner,mosaic,tcp_finger}. We employ and compare two approaches using HTTP and/or IP headers.    

\heading{(1) App Extraction using App Strings:} Although HTTPS dominates the general web connections, the usage of HTTP is still significant in mobile apps according to recent research ~\cite{inconsistent,networkframe,locationtracking,ample,lookat,bugfix,appcracker}. Although Android defaults to use HTTPS traffic in all apps since 2018, it is not strictly enforced, and still allows developers to change configuration to use HTTP~\cite{androidP}. Indeed, it has been observed in recent study that only part of communication (e.g., initial requests) are secured over HTTPS~\cite{debate,mixcontent,appcracker}. This may be due to a few reasons. First, advertisements (ads) traffic accounts for a significant portion in mobile networks and ads traffic is mostly carried over HTTP ~\cite{locationtracking,breaking,mixcontent,bugfix,darkside}. Further, HTTPS adds significant costs (e.g., significantly increasing latency and energy consumption) due to cryptographic operations and a required extra handshake, which is critical in mobile networks~\cite{httpscost,appcracker}.  For the same reason, it has also been observed in previous study that most malicious traffic is carried over HTTP ~\cite{impending,nazca,dissecting}.

 \begin{figure}[tb]
\begin{center}
\parbox{0.5\textwidth}{
  $GET /open/confirm.htm?pkg= com.sina.news$
\caption{A HTTP header with an app string}\label{pkt}}

\end{center}
\end{figure} 
This approach thus focuses on HTTP traffic and extract app information revealed in HTTP header. That is, we extract the IP packets containing the app string field in the header. The app string often contains the name of the app binary file. 10\% of our traffic includes explicit app strings. An example of packets with an app string is presented in Fig.~\ref{pkt}. We assume each unique app string as an app. By doing so, we gather 5870 app strings from our dataset.

\heading{(2) App Extraction using IP:} 
Most mobile apps require network connection between devices and their host servers while being downloaded, installed, or executed. One way to extract app information without explicit app strings (e.g., HTTPS traffic) is thus that we use each or a group of destination IPs as the counterpart of apps since an IP may represent a server hosting a specific app. As a first step to deal with traffic without app strings, we explore a naive approach to treat an IP as an app. Note that a single IP often does not reliably represent an app for several reasons~\cite{robust,appscanner}. Indeed, in Section~\ref{accuracy}, we will show identifying compromised devices only using destination IPs results in high false positive rates. Meanwhile, we discuss how it can be improved in Section \ref{discussion}.  To fairly compare two approaches, we use the same HTTP dataset with which 6150 destination IPs are extracted.
\begin{figure}[htbp]
\begin{center}
\parbox{0.5\textwidth}{
\centering 
\subfigure[The CDF of the number of devices for all apps]{
 \epsfig{file=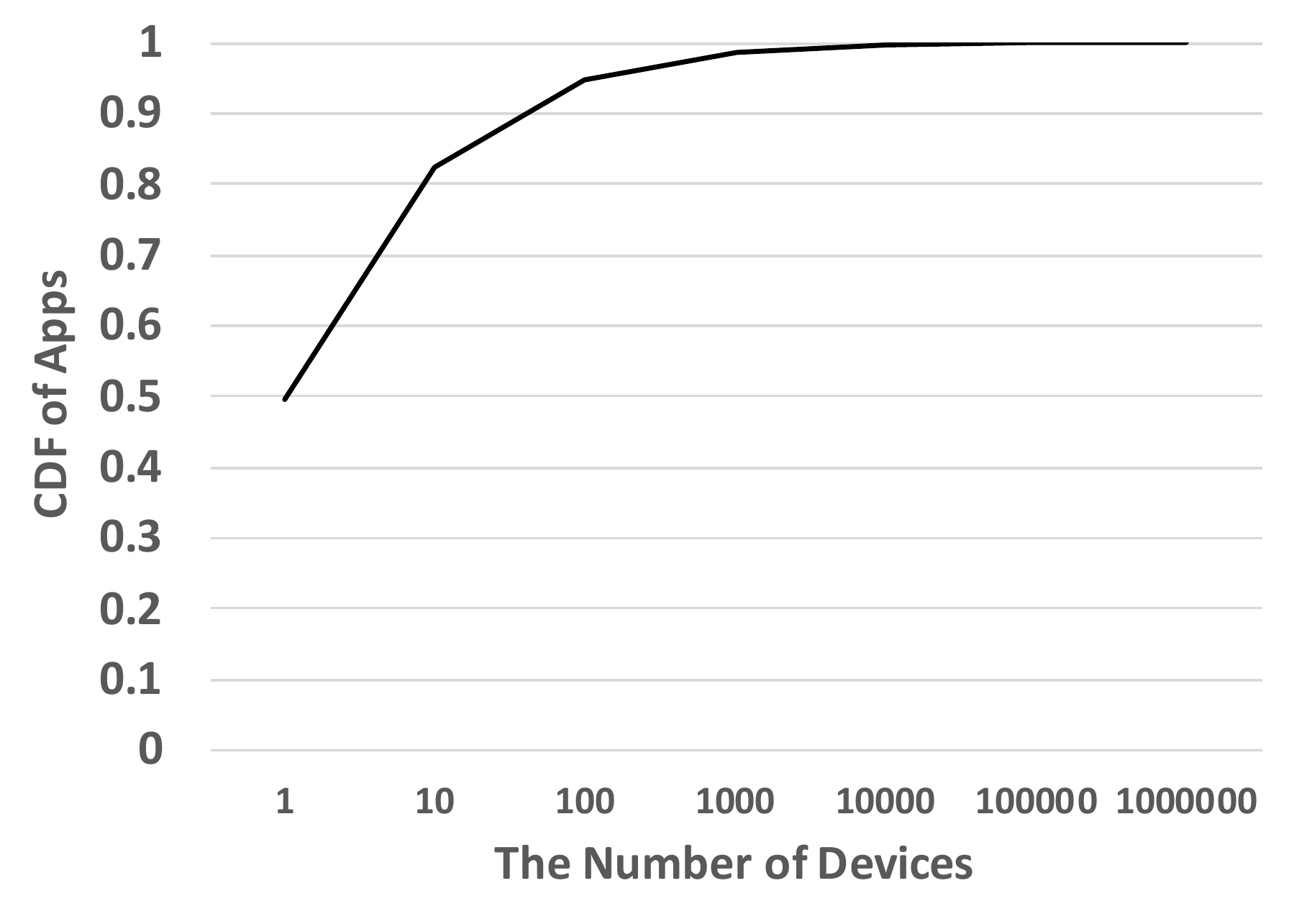,width=0.24\textwidth, height=0.15\textheight}\label{num_devices_app}}
\subfigure[The CDF of the number of devices for all destination IPs]{
\epsfig{file=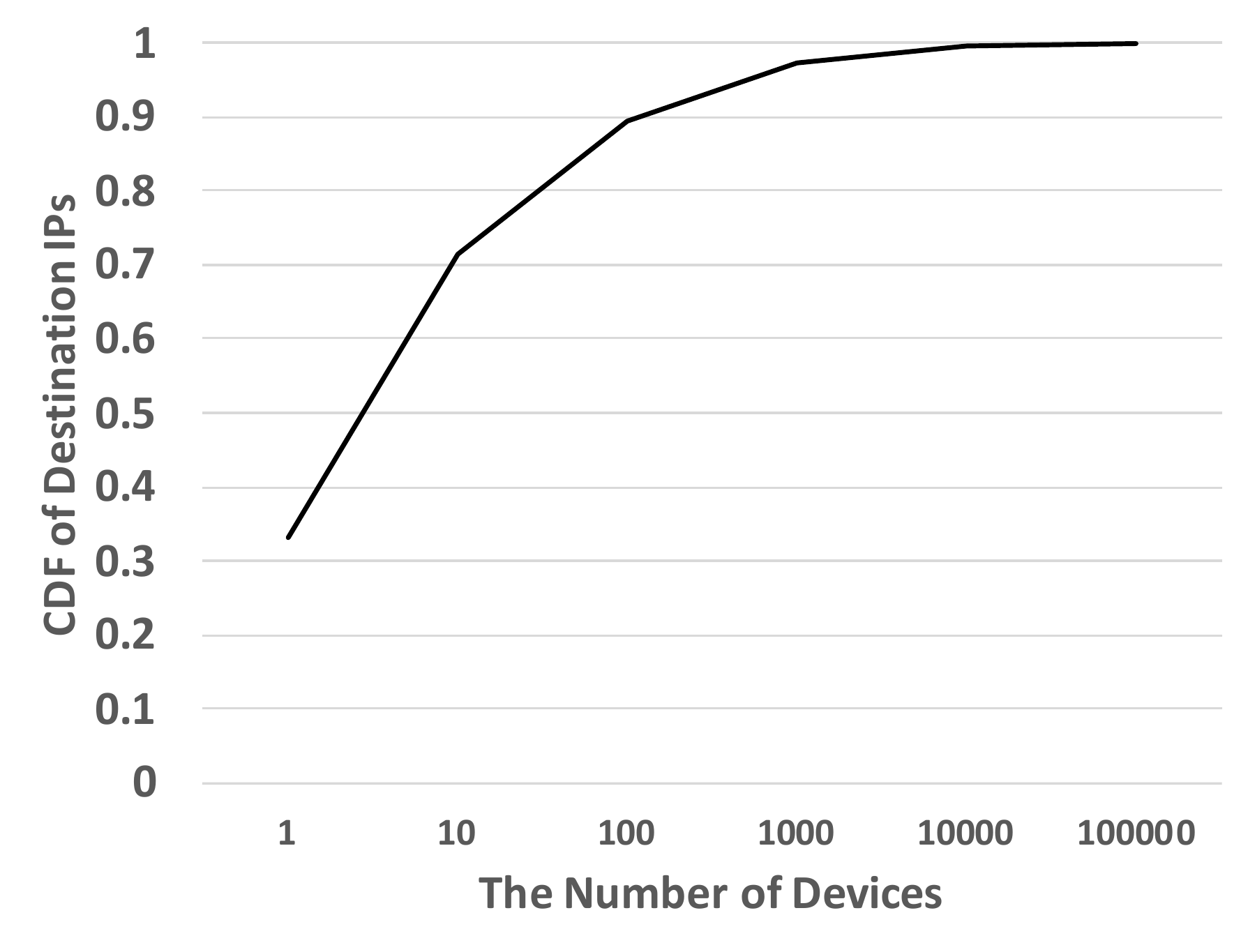,width=0.24\textwidth, height=0.15\textheight}\label{num_devices_ip}}
}%\vspace{-.17in}
\caption{The distribution of number of devices}\label{cdf_num_devices}

\end{center}
\end{figure}

\subsubsection{A Device-App Graph}\label{sec:device_cnt}

By the end of this process, we have a mapping between devices and apps (either app strings or destination IPs), i.e, edges. Fig.\ref{num_devices_app} presents the CDF of the number of devices where the x-axis represents the number of devices having each app string and the y-axis represents the corresponding CDF (i.e., the portion of apps). Note that nearly 50\% of apps are having only one device. This is mainly because app strings may also include version names or market names of the apps such as com.sina.news-7.19.3 and com.supercell.clashofclans.baidu, which we consider as individual apps.

Fig.\ref{num_devices_ip} presents the CDF of the number of devices where the x-axis represents the number of devices connecting to each destination IP and the y-axis represents the corresponding CDF (i.e., the portion of destination IPs). Fig.~\ref{cdf_num_devices} suggests that each device will have relatively more shared IPs with another, compared to the number of shared apps. We discuss the effect of this distribution in Section~\ref{accuracy}.

\subsection{Ground Truth Sets and Definitions for Detecting Compromised Devices}

Our approach requires the small sets of ground truth about compromised and not-compromised devices to apply BP. However, given the vast number of unknown apps and little, if any, knowledge on devices in practice, it is often hard to build a ground truth set. We first collect ground truth sets for apps (destination IPs and app binaries) (Section~\ref{sec:app_groundtruth}) and construct ground truth sets for devices (Section~\ref{sec:device_groundtruth}).

\subsubsection{Ground Truth Sets for Apps}\label{sec:app_groundtruth}

One may utilize any one or multiple intelligence sources to build a ground truth set for apps. We use VirusTotal (VT)~\cite{virustotal} to collect a ground truth set for apps. VT is a security intelligence portal for IPs, URLs and binaries, based on third-party anti-virus engines, widely used in the literature for building ground truth~\cite{beyondgoogleplay,unknown_malice,mastino_mining,impending}. For each query, VT aggregates the responses from more than 50 engines, each of which categorizes the queried IP, binary or URL to \emph{malicious} or \emph{benign}. %It is important to note that the source of establishing ground truth is independent of our approach.

\begin{figure}[tb]
\begin{center}
\parbox{0.4\textwidth}{
\centering
    \epsfig{file=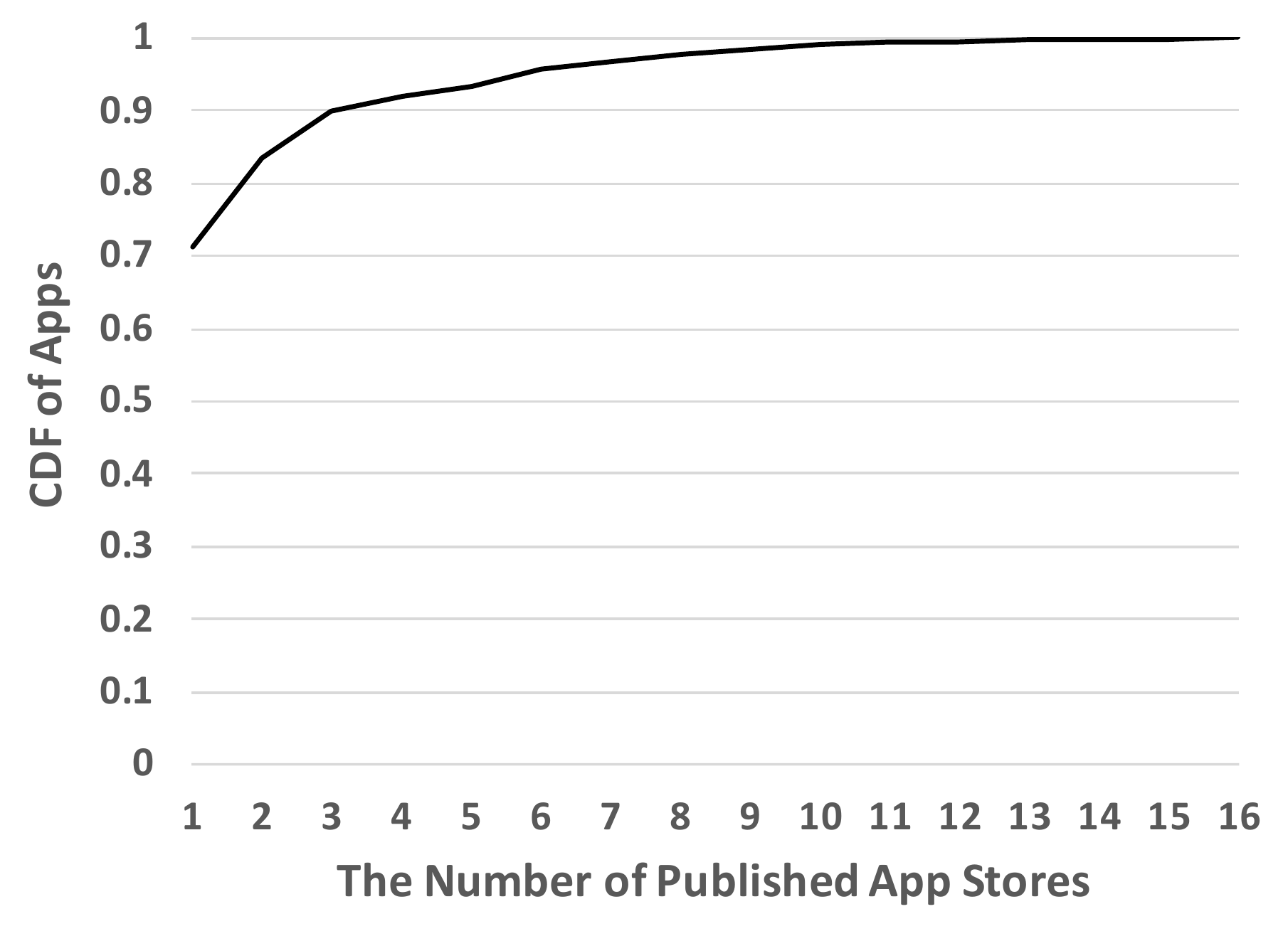,width=0.35\textwidth, height=0.18\textheight}
\caption{The CDF of the number of published app stores}\label{market_stat}}

\end{center}
\end{figure}
\heading{(1) App Binary Analysis}: To build a ground truth set, we first attempt to download android binaries from 16 popular Chinese app stores ~\cite{beyondgoogleplay} by searching the app strings extracted in Section ~\ref{sec:app_extraction}. App string in the traffic is sometimes not readable, as it could be truncated or represented as simple digits or a serial number~\cite{appprint,ample}. Among 5870, 2367 app strings were found in at least one of 16 app stores. Fig.\ref{market_stat} presents the Cumulative Distribution Function (CDF) of the number of app stores where the x-axis represents the number of app stores having each app string and the y-axis represents the corresponding CDF (i.e., the portion of apps). Note that we also verify that the binaries are indeed generating the corresponding app strings by capturing network traffic while installing and executing binaries on two mobile devices (i.e., Samsung Galaxy Note4 and Sony Xperia Z3 Dual).

We upload the binaries to VT and check whether it is marked as malicious. Note that 29\% of app strings are published in multiple app stores, as shown in Fig.\ref{market_stat}. However, we observe that the maliciousness of each app is the same regardless of the app stores where it is downloaded. This observation  agrees with that from prior research: the app string often correctly represents a specific app ~\cite{beyondgoogleplay,usagepattern}. Furthermore, Haoyu \emph{et al.} observed that recent mobile malware does not spread by repackaging as much as before in a large-scale study on android apps ~\cite{beyondgoogleplay}. We thus argue that it is reasonable to rely on the app string to identify and evaluate each app.

Among 2367 binaries, 1711 apps are flagged as malicious by at least one VT engine and 656 apps are not flagged by any engine. Previous research suggests that evaluation based on VT may have a limited coverage or noise due to multiple reasons ~\cite{appriskanalysis,beyondgoogleplay,unknown_malice,droppereffect}. To reduce potential false positives, we label each app using thresholds as follows.

If an app is detected as malicious by more than or equal to $vt$ number of engines among the 60 VT engines, we label the app as \textbf{bad}; if an app is detected  as malicious by less than $vt$ engines, we label the app as \textbf{suspicious}; if an app is not detected as malicious by any engine, we assume that the app is \textbf{good}; if we are not able to find corresponding binaries, we consider the app as \textbf{no-info}.\footnote{For consistency reasons, we follow previous work~\cite{polonium,guiltfile,fraudeagle,speagle} in choosing ``good'', and ``bad'' as BP labels though they may not sound technical.}. 

Note that we aim to identify unknown compromised devices that may have installed unknown malicious apps. Many app stores are known to perform a vetting process to identify and remove malicious apps from the stores ~\cite{beyondgoogleplay,unknown_malice,getoff}. As it is relatively easy to detect popular yet bad apps through such a general vetting process,
we exclude popular apps and app libraries. Concretely, we consider an app popular if it is used by more than $N_{p}$ number of devices. This filtering is important as it helps us avoid a  number of false positives which can be induced by false association in a graph-based approach~\cite{snare,maldomain,codaspy_issa,exposure}. We shall discuss the effect and limitation caused by this filtering in Section~\ref{accuracy} and Section~\ref{discussion}.

Table~\ref{gt_app} summarizes the number of bad apps with various $vt$ and $N_p$ thresholds.
\begin{table}[tb]%%
\begin{center}
\small
\begin{tabular}{|c||c|c|c|c|c|}
\hline
\backslashbox[2.5cm]{App\\popularity}{VT}
 &  $vt=3$  & $vt=4$  &$vt=5$  &$vt=6$  &$vt=7$ \\
\hline\hline
Not-filtered &1195&1060&955 & 845&764\\
\hline
$N_{p}$ = 10000 &1190&1056&951 & 841&761\\
\hline
$N_{p}$ = 5000 &1189&1056&951 & 841&761\\
\hline
$N_{p}$ = 1000 &1178&1047&943 & 833&753\\
\hline
\end{tabular}

\caption{The number of bad apps with varying $vt$ and $N_p$ thresholds}
\label{gt_app}

\end{center}
\end{table}

\heading{(2) Destination IP Analysis}: Towards dealing with traffic without explicit app strings, we inspect destination IPs that devices are connecting to. Particularly, we checked 6150 destination IPs in our dataset extracted in Section~\ref{sec:app_extraction} against VT. Following previous research~\cite{codaspy_issa}, we label an IP  as \textbf{bad}, if the IP is detected as malicious by two or more engines in VT; if the IP is detected  as malicious by only one engine, we label it as \textbf{suspicious}.  By doing so, we find 528 bad IPs.

%\mashael{MS: Also clarify why those could cause a large false positive.}

% %observed that only 38.3% of these malware samples are repackaged apps. This result suggests that app repackaging is no longer the main way for malware spreading. 

\subsubsection{Ground Truth Sets for Devices}\label{sec:device_groundtruth}

Given the ground truth set for apps, we define a \textbf{bad device} as one using more than or equal to $N(A_b)$ number of bad apps, where $A_b$ is the set of bad apps; we define a \textbf{good device} as one not using any bad and suspicious apps. Note that we have two ground truth sets for apps (i.e., based on (1) app string and (2) destination IP). Accordingly, we also build two ground truth sets for devices.

\heading{(1) An App Binary Based Ground Truth Set: } Table~\ref{gt_device} summarizes the number of devices given the ground truth sets in Table.~\ref{gt_app}.

\begin{table}[tb]%%
\begin{center}
\small
\begin{tabular}{|c|p{0.58cm}|p{0.58cm}|p{0.58cm}|p{0.58cm}|p{0.55cm}||p{0.65cm}|}
\hline
\backslashbox[2.1cm]{App\\popularity}{VT}
 &  $vt=3$  & $vt=4$  &$vt=5$  &$vt=6$  &$vt=7$ & good device  \\
\hline\hline
Not-filtered &25847&16889&15449 & 14391&9280&53162\\
\hline
$N_{p}$ = 10000 &7547&5347&4385 & 3621&3169&25782\\
\hline
$N_{p}$ = 5000 &6871&5347&4385 & 3621&3169&24231\\
\hline
$N_{p}$ = 1000 &2759&2371&2004 & 1400&1086&12923\\
\hline
\end{tabular}

\caption{The number of devices with varying $vt$ and $N_p$ thresholds}
\label{gt_device}

\end{center}
\end{table}

\heading{(2) A Destination IP-based Ground Truth Set: }
We build a ground truth set for devices given the ground truth sets in Section~\ref{sec:app_groundtruth}(2).
By doing so, we found 442 bad devices connecting to 2 or more bad IPs and 13794 good devices.

\section{Experimental Results and Analysis}\label{sec_analysis}

%We empirically verify our hypothesis that (1) there exist homophily relationships between devices and apps installed in the devices; (2) a graph-inference approach based on such association can correctly classify instances into two distinct classes: compromised devices and not-compromised devices.

\subsection{Experimental Setup}

\begin{table}[tb]%%
\small
\centering
%\begin{center}
\begin{tabular}{|p{1.2cm}|p{5.5cm}|p{1cm}|}\hline
\textbf{Notation}  & \textbf{Description} & \textbf{Default}\\
\hline
$N_p$        & The threshold to define a popular app  & 1000 \\
\hline
$vt$        & The threshold for the number of VT engines detecting the app as malicious &  5 \\
\hline
$N(A_B)$        & The number of bad apps the bad device has & 2\\
\hline
$\epsilon$ & The edge potential parameter & 0.51\\
\hline
$\delta$ & The initial belief parameter & 0.99\\
\hline
\end{tabular}
\caption{List of parameters for experiments and default values}

\label{param}
%\end{center}
\end{table}

Table \ref{param} summarizes the notation for the parameters used in our experiments.

\textbf{\emph{BP Implementation:}} We implemented BP in C following the implementation in~\cite{maldomain}, as it was shown to be fast and scalable for large graphs. It only takes 1.25 seconds to run 10 BP iterations on average for our graphs.

\textbf{\emph{Ground Truth Sets for BP:}} In our bipartite graph, we have two types of nodes: apps and devices. Although BP can be used to classify both apps and devices in principle, we focus on the classification of devices. We thus consider the badness/goodness of apps as unknown. %Considering apps as unknown eliminates the possibility of the introduction of biases in the device ground truth. \ting{I don't quite understand this. what bias does it refer to? Why will there be bias?}
%Nabeel: I added the above sentence to address the review comment about device ground truth and then doing BP over them again.
Hence, BP inference is driven by two different sets of device ground truth labels: bad devices ($D_B$) and good devices ($D_G$). The number of instances in each set is described in Section~\ref{sec:device_groundtruth}. Note that the original data set is not balanced.
It is natural, however, that unbalanced initial belief leads to a biased set dominating the final result~\cite{snare,maldomain}. We thus randomly choose an equal number of instances from each set to avoid any such bias. For example,  as described in Table~\ref{gt_device}, there are 12923 good devices and 2004 bad devices if we set $vt=5$ and  $N_{p}= 1000$. In such a case, we use all of 2004 bad devices as $D_B$, and randomly choose 2004 out of the 12923 good devices as $D_G$.

\textbf{\emph{Cross Validation:}} To validate our approach, we perform k-fold cross validation. We randomly divide each of $D_B$ and $D_G$ into k folds and run BP $k$ times. In each BP run, we use one of the $k$ folds as a testing set and the remaining $k-1$ folds as a training set. We rotate the testing fold across the $k$ folds in the $k$ runs of BP and the final results are the average of the results from the $k$ BP runs. For our data set, BP converges fast, and therefore we do not limit its number of iterations. 
%use $k$-1 folds as training set, and use the remaining fold as a testing set.  In our observations, varying $k$ results in no significant difference in performance, so we next report the 10-fold classification results.

As discussed in Section~\ref{sec_bp}, each node has two belief scores representing its badness ($P(Bad)$) and goodness ($P(Good) = 1 - P(Bad)$), respectively. For simplicity, we only mention the badness scores in this section. In each BP run, we consider the devices in the testing set as unknowns and hence, their initial beliefs are set to 0.5, as described in Table~\ref{bp_table}(a). The initial beliefs of devices in the training set are set according to their ground truth labels. Specifically, we set $\delta$ = 0.99 
and hence, the initial badness beliefs of devices in the training set from $D_B$ are set to 0.99, and 0.01 for those in the training set from $D_G$. 
%so that the badness of devices in the training set extracted from $D_B$ will be set as 0.99, and that of devices in the training set extracted from  $D_G$ will be set as 0.01. Note that we vary $\delta$ within the range from above 0.5 to below 1, resulting in no significant difference in performance, as long as $\delta$ reflects its badness and goodness of ground truth labels correctly. Given the training sets, we run BP  for each fold until it converges. In our dataset, BP quickly converged, so we do not limit the number of iterations for BP.  

Devices in the testing set are labeled based on their average final beliefs. Specifically, we vary the threshold for final beliefs, and classify a device whose final belief is above the threshold as bad. Otherwise we classify it as good. We then compute the true positive rate as the number of bad devices that are correctly classified to the total number of bad devices in the test set. Similarly, the false positive rate is computed as the number of good device that are misclassified to the total number of good devices in the test set.
%After BP converges, we label devices in the testing set based on the final belief scores, based on which we measure true positive rate and false positive rate. 
%Specifically, we vary the threshold for final belief scores, and consider a device whose final belief score is above the threshold as bad; otherwise consider a device as good. In our context, true positive means that a bad device is correctly classified as bad; false positive means that a good device is incorrectly classified as bad. 

\subsection{Device Classification Accuracy}\label{accuracy}

To show the detection accuracy, we present a series of ROC curves where: the x-axis represents the false positive rate (FPR), the y-axis represents the true positive rate (TPR), and each point in ROC curves represents different threshold for BP final belief scores. While doing so, we vary the parameters in Table \ref{param}, that might have effect on the performance.

\begin{figure*}[htbp]
\begin{center}
\parbox{1.0\textwidth}{
\centering 
\subfigure[Varying ground truth sets]{
\psfig{file=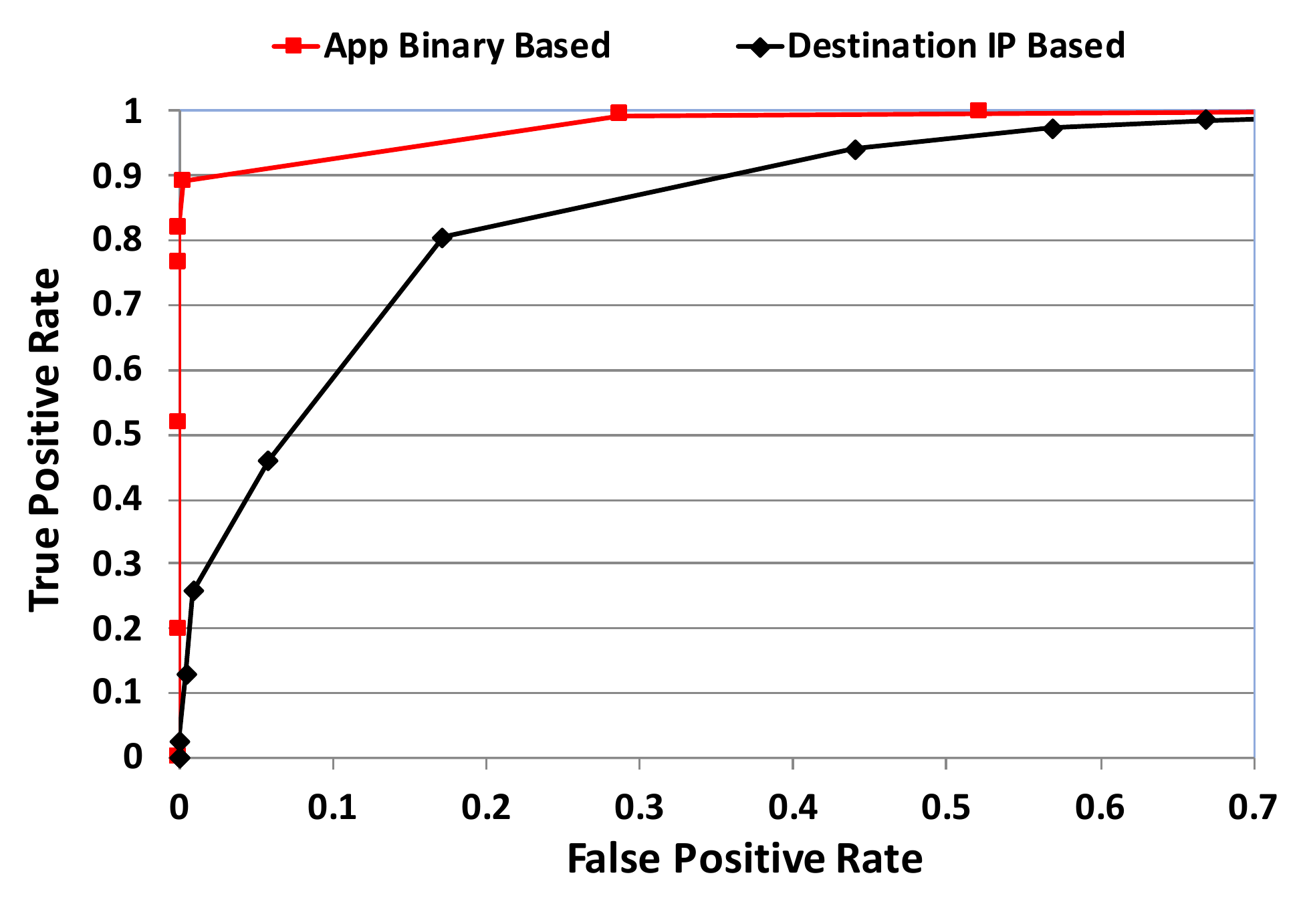,width=0.23\textwidth, height=0.18\textheight}\label{device_ip_app_roc}}
\subfigure[Varying popular app definitions]{
\psfig{file=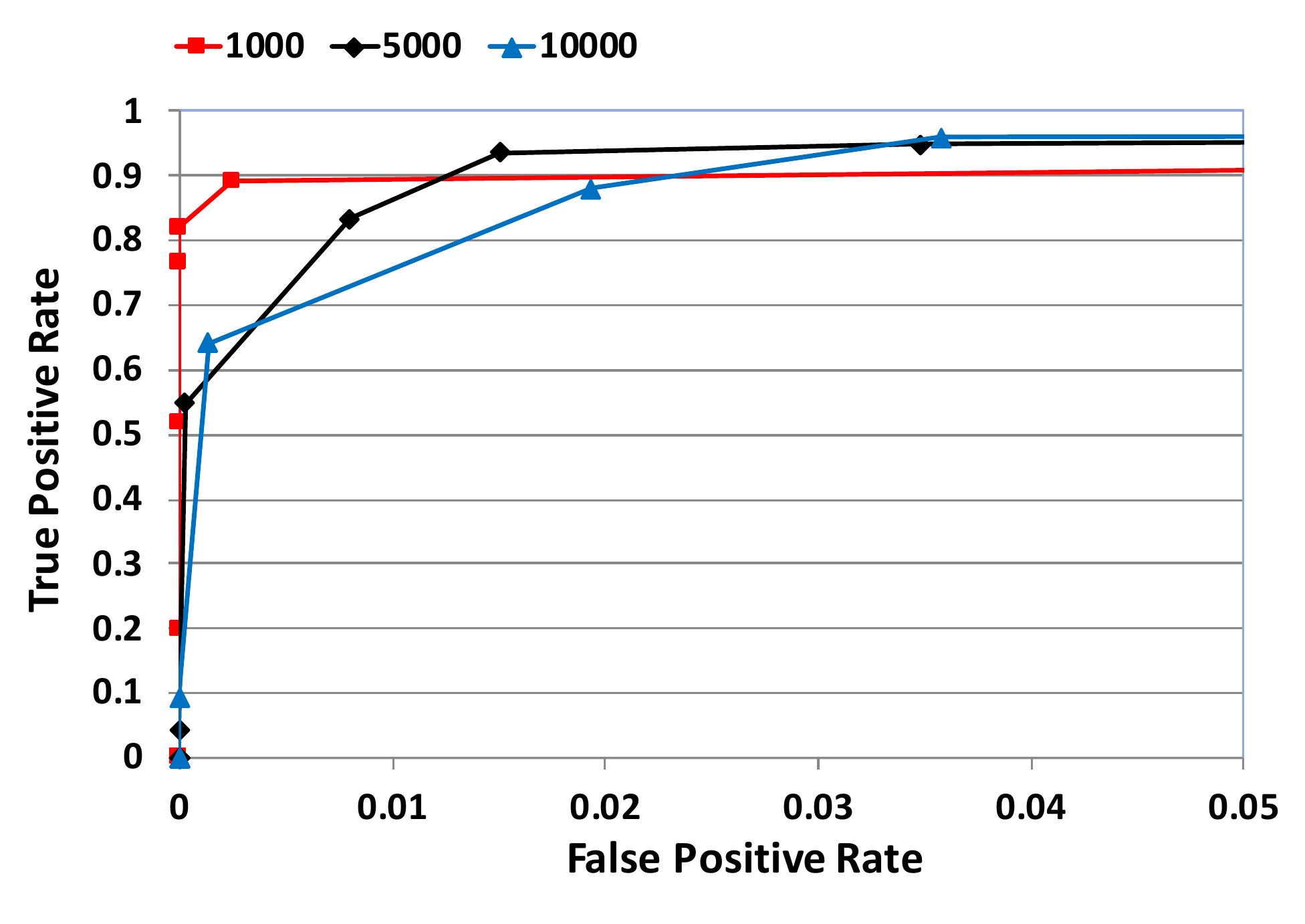,width=0.23\textwidth, height=0.18\textheight}\label{roc_popularapp}}
\subfigure[Various $vt$]{
\psfig{file=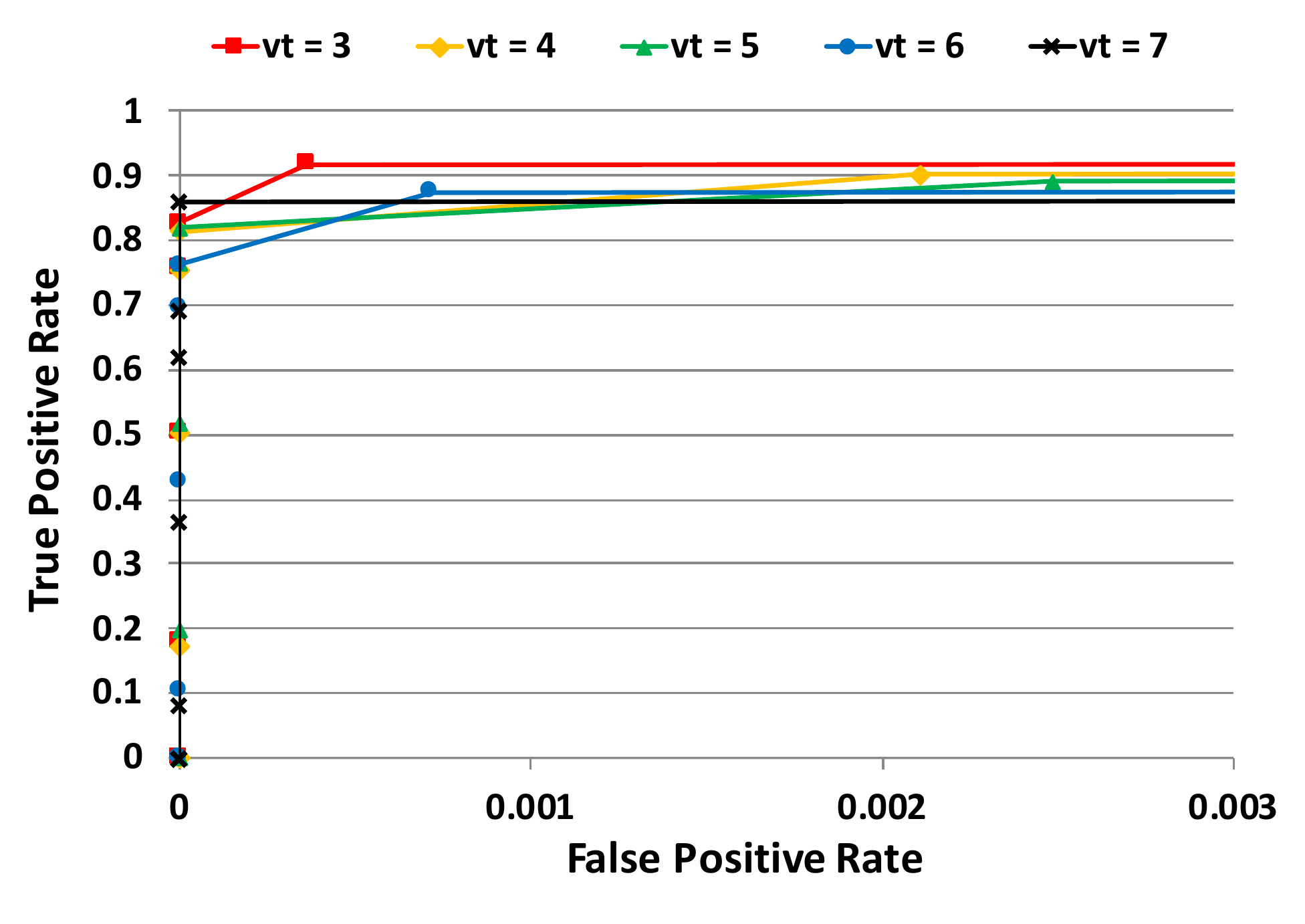,width=0.23\textwidth, height=0.18\textheight}\label{vt5_3_1}}
\subfigure[Various $\epsilon$]{
\psfig{file=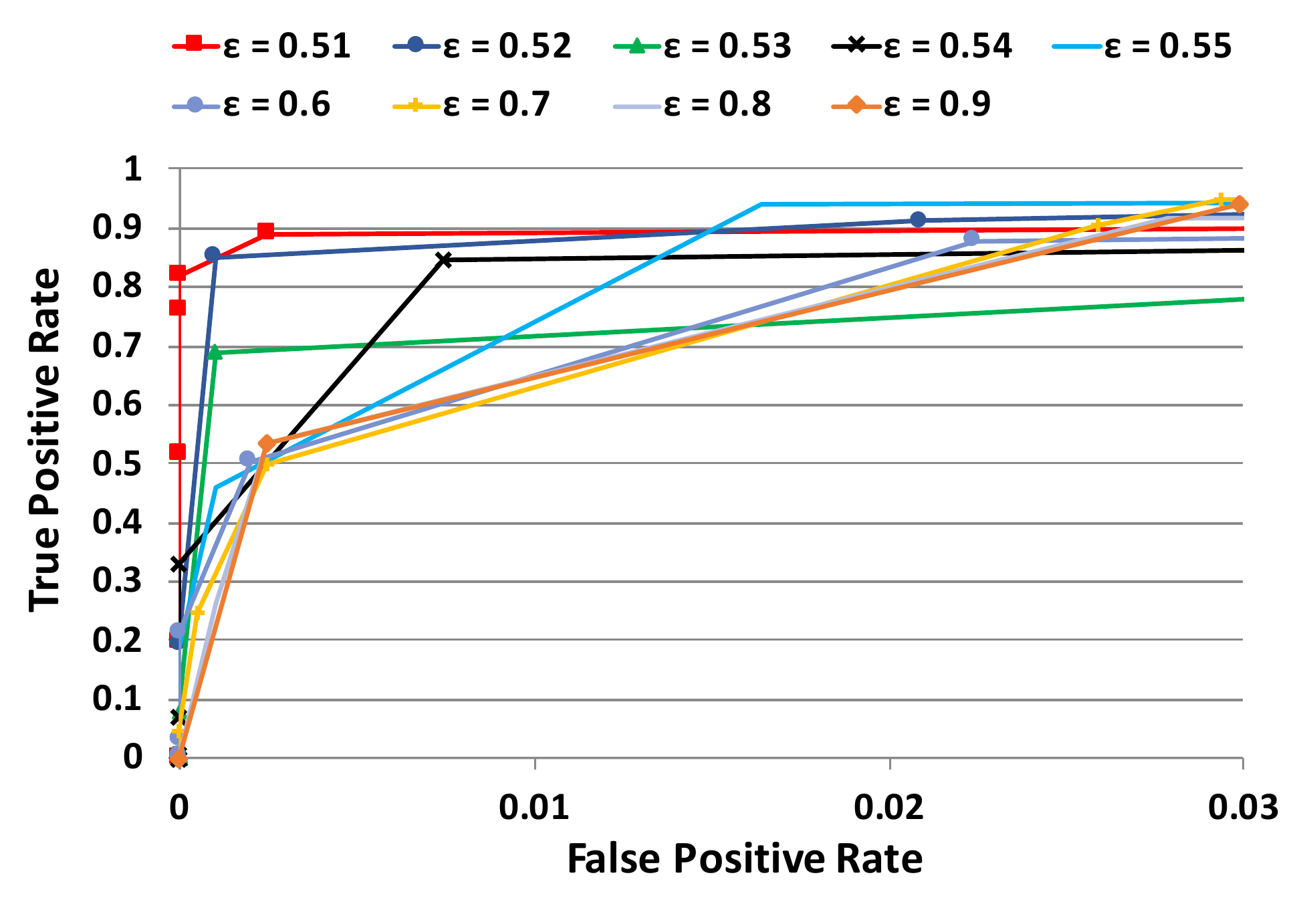,width=0.23\textwidth, height=0.18\textheight}\label{roc_epsilon}}
}
\caption{ROC curves with various parameter setting}\label{all_roc_curves}
\end{center}

\end{figure*}

\heading{Varying the Ground Truth Set.} Fig.\ref{device_ip_app_roc} shows the ROC curves while varying ground truth sets built based on destination IPs and app binaries. The figure clearly shows that the classification with the destination IP based ground truth set provides modest accuracy. The best FPR and TPR it could achieve are 0.17 and 0.804, respectively. Such an accuracy is not acceptable in practice as it misclassifies a considerable number of devices. This is mainly because a single destination IP does correctly represent an app and thus a user's general behavior, while most activities on mobile devices incur through apps. Hence, if we treat each IP as an app, our approach will consider all devices connect to the same IP as related; which in turn results in  a lot of false associations harming the accuracy of graph-inference approach. In fact, 39\% IPs are used for servers hosting two or more unrelated apps in the given IP. In the following, we thus focus on detection with an app-binary based ground truth set, and provide discussion about how we can improve detection with a destination IP-based ground truth set in Section~\ref{discussion}.

\heading{Varying the Definition of Popular Apps.} Fig.\ref{roc_popularapp} shows the ROC curves with varying the definition of popular apps.  Although not significant, the figure shows that with less filtering such as $N_P= 10000$, the false positive increases. This is expected because many devices, if not all, using popular apps will be considered related, resulting in false associations~\cite{snare,maldomain,codaspy_issa,exposure}.

\heading{Varying $vt$ Values.} Note that there is no consensus on the right value for $vt$ ~\cite{appriskanalysis,beyondgoogleplay}. We thus show the results with various $vt$, i.e., $vt = {3, 4, 5, 6, 7}$ in Fig.~\ref{vt5_3_1}. As the figure shows, there is no significant difference on the false positive rates and true positive rates while changing $vt$. We thus use $vt=5$ for further discussion in the following without loss of generality.

\heading{Varying Edge Potentials $\epsilon$.} Fig.\ref{roc_epsilon} shows the ROC curves  while varying $\epsilon$. To be specific, when $\epsilon = 0.51$, we can achieve 0.89 TPR with only 0.002 FPR. Interestingly (as it is different from BP behavior in other applications such as ~\cite{codaspy_issa,maldomain}), our results are sensitive to $\epsilon$, an edge potential parameter of BP. As shown in Fig.\ref{roc_epsilon}, as $\epsilon$ increases, the false positive rates increase. Furthermore, we achieve the best accuracy with low values of $\epsilon$ (e.g., 0.51). In the next section, we present an in-depth analysis to demonstrate distinctive characteristics of the mobile traffic data that lead to this behavior. 

\subsection{In-Depth Analysis of BP behavior}\label{sec_graphstruct}

 Identifying the most effective edge potential values to accurately classify graph nodes using BP is a known problem, and recent work aims to automate the process of identifying such values\cite{edgepotential}. That said, it has also been observed in a variety of previous work \cite{maldomain,codaspy_issa,fraudeagle} that the accuracy of BP is not significantly impacted by different $\epsilon$ values. On the contrary, recall that our results are sensitive to $\epsilon$ as discussed in Section~\ref{accuracy}.  In this section, we shed light on why edge potential value  $\epsilon$ has an impact on our results, by providing in-depth analysis on distinctive network properties of two bipartite graphs from different applications. In the first graph (\emph{Mobile}), $\epsilon$ has obvious impact on accuracy, while in the other (\emph{DNS}), $\epsilon$ has no notable impact on accuracy. %We discuss two ground-truth label-annotated set. In each set, we have three types of nodes: bad, good, and unknown.  

\emph{Mobile} represents the bipartite graph built from our dataset. For our experiments, we have used various ground truth sets while changing $vt$ to define a bad device, which,  
as shown in Fig.\ref{vt5_3_1}, have no significant impact on false positive rates and true positive rates. We also note that nodes in the ground truth drawn with different $vt$s do not have much topological difference. Without loss of generality, we thus use the ground truth drawn with $vt=5$ to provide analysis in the following.

 \emph{DNS} represents the bipartite graph between domains and IPs built from the active DNS dataset in \cite{codaspy_issa}. To compare the impact of $\epsilon$ in different networks, we obtained the domain-ip bipartite graph and the ground truth labels on domains used in their work~\cite{codaspy_issa}. Concretely, in their bipartite graph, domains and IPs are connected as edges, each of which represents a domain resolving to an IP. They collected ground truth for bad domains by checking domains against VT and for good domains by checking domains against Alexa top list~\cite{alexa}. 

\begin{figure}[tb]
\begin{center}
\parbox{0.4\textwidth}{
\centering 
\epsfig{file=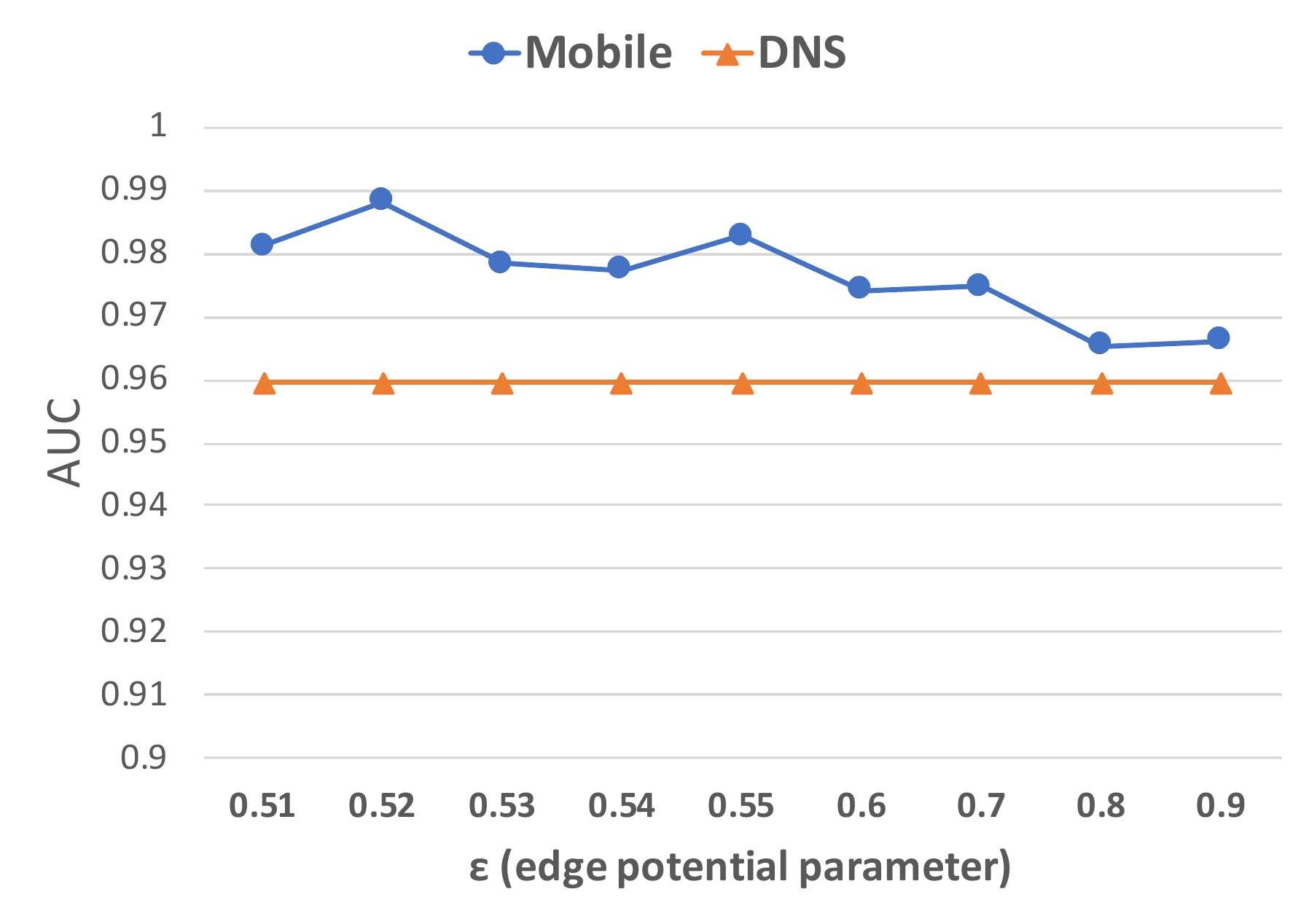,width=0.38\textwidth, height=0.2\textheight}
%\vspace{-.21in}
}
\caption{AUC when varying $\epsilon$ for Mobile and DNS}\label{auc}

\end{center}
\end{figure}
To clearly  capture the sensitivity to $\epsilon$ in each of the two graphs (\emph{Mobile} and \emph{DNS}), we measure the area under the ROC curve (AUC). Fig.\ref{auc} shows AUC with varying $\epsilon$,  where the x-axis represents $\epsilon$ and the y-axis represents the corresponding AUC for each graph. The figure clearly shows that the classification accuracy in \emph{Mobile} gets lower (from 0.98 to 0.97), as we increase $\epsilon$ by 0.1 (0.51, 0.6, 0.7, 0.8, 0.9). On the other hand, the classification accuracy in \emph{DNS} stays almost the same (0.96), regardless of $\epsilon$. We argue that this different behavior of BP is due to the network structures and the topological locations of nodes in the ground truth. 

For any two nodes $S$ and $T$ in the graph, their impact on each other depends on multiple variables, the most important of which are: (1) the length of the path between  $S$ and $T$, (2) the number of paths between $S$ and $T$, and (3) the edge potential parameter $\epsilon$.
%and how much $S$'s  badness score $b_S$  affects $T$'s  badness score $b_T$ during BP relying on their topological location.  According to message and belief score computation functions given by \ref{eq:msg} and \ref{eq:belief} in Section \ref{sec_bp}, there are three factors which  affect $b_S$'s impact on $b_T$ : 
\begin{comment}
\begin{itemize}
    \item the length of the path between  $S$ and $T$,
    \item the number of paths between $S$ and $T$, and 
    \item the edge potential parameter $\epsilon$.
\end{itemize}
\end{comment}
%(1) the length of the path between  $S$ and $T$; (2) the number of paths between $S$ and $T$; (3) the edge potential parameter $\epsilon$.  
First, the longer the path between $S$ and $T$, the smaller $S$'s impact on $T$. This is because the edge potential diminishes as it travels on the path between the two nodes (due to fraction multiplications as many as the length of the path.) %More importantly, if the path between $S$ and $T$ is relatively long (e.g., 20 as opposed to 2), the impact will greatly diminish  irrespective of  $\epsilon$. 
As a result, the final badness score will be insensitive to $\epsilon$ in case of graphs with longer paths.

Second, the larger the number of paths between $S$ and $T$, the higher the impact of $S$ on $T$. This is because the final belief at $T$ is a function of the product of messages received on each path from $S$ to $T$. For example, assume that a bad node $S$ has $p$ paths to $T$, then $S$ sends a bad message $m_B(i)$ and a good message $m_G(i)$ on a path $i$. Since $S$ is bad, $m_B(i)$ is larger than $m_G(i)$. The final bad (good) impact of $S$ on $T$ is a function of the product of the $m_B(i)$ ($m_G(i)$) messages from all the $p$ paths. The larger the number of paths ($p$), the higher the difference between the $m_B(i)$ product and the $m_G(i)$ product, and hence, the higher the final badness score (due to the assumption that $S$ is bad in the example). 
%Second, the larger the number of paths between $S$ and $T$, the higher the impact of $S$ on $T$. This is because the final belief at $T$ is a function of the product of messages received on each path from $S$ to $T$. For example, assume that a bad node $S$ sends a bad message $m_S(B)$ and a good message $m_S(G)$ on a path $p$. Since $S$ is bad, $m_S(B)$ is larger than $m_S(G)$. . The larger the number of paths between $S$ and $T$ is, the higher the difference between $m_S(B)$ and $m_S(G)$ is; hence, the higher the final badness score is. 
%each message  be multiplied as many as the number of paths. To be specific, assume that bad node $S$ sends a bad message $m_S(B)$ and a good message $m_S(G)$ on a path $p$. Since $S$ is bad, $m_S(B)$ is larger than $m_S(G)$. Note that the final belief score is a function of the product of $m_S(B)$/$m_S(G)$. The larger the number of paths between $S$ and $T$ is, the higher the difference between $m_S(B)$ and $m_S(G)$ is; hence, the higher final badness score is.

%This is because the badness and goodness scores ($\epsilon$ or $1-\epsilon$) are multiplied as many as the number of paths.

Finally, if we set $\epsilon = 1$, the path length will no longer have any impact, because length-1 has the same impact as length-1000; if we set $\epsilon$ close to 0.5, $b_S$'s impact on $b_T$ greatly diminishes except for very short paths (e.g., 2). \\
We next compare two datasets from the two graphs ({\em Mobile} and {\em DNS}) in terms of there topological features. Specifically, we are interested in nodes in the ground truth set.

\textbf{Shortest path length:} Consider two clusters: bad ($C_B$) and good ($C_G$). The important intuition behind BP using homophily relationship is that each cluster's intra-cluster distance is supposed to be low, whereas inter-cluster distance between two clusters is supposed to be high. We thus  measure the intra-cluster and inter-cluster distances in terms of the shortest path lengths between all pair of nodes in $C_B$ and $C_G$.

\begin{figure}[htbp]

\begin{center}
\parbox{0.4\textwidth}{
\centering
    \epsfig{file=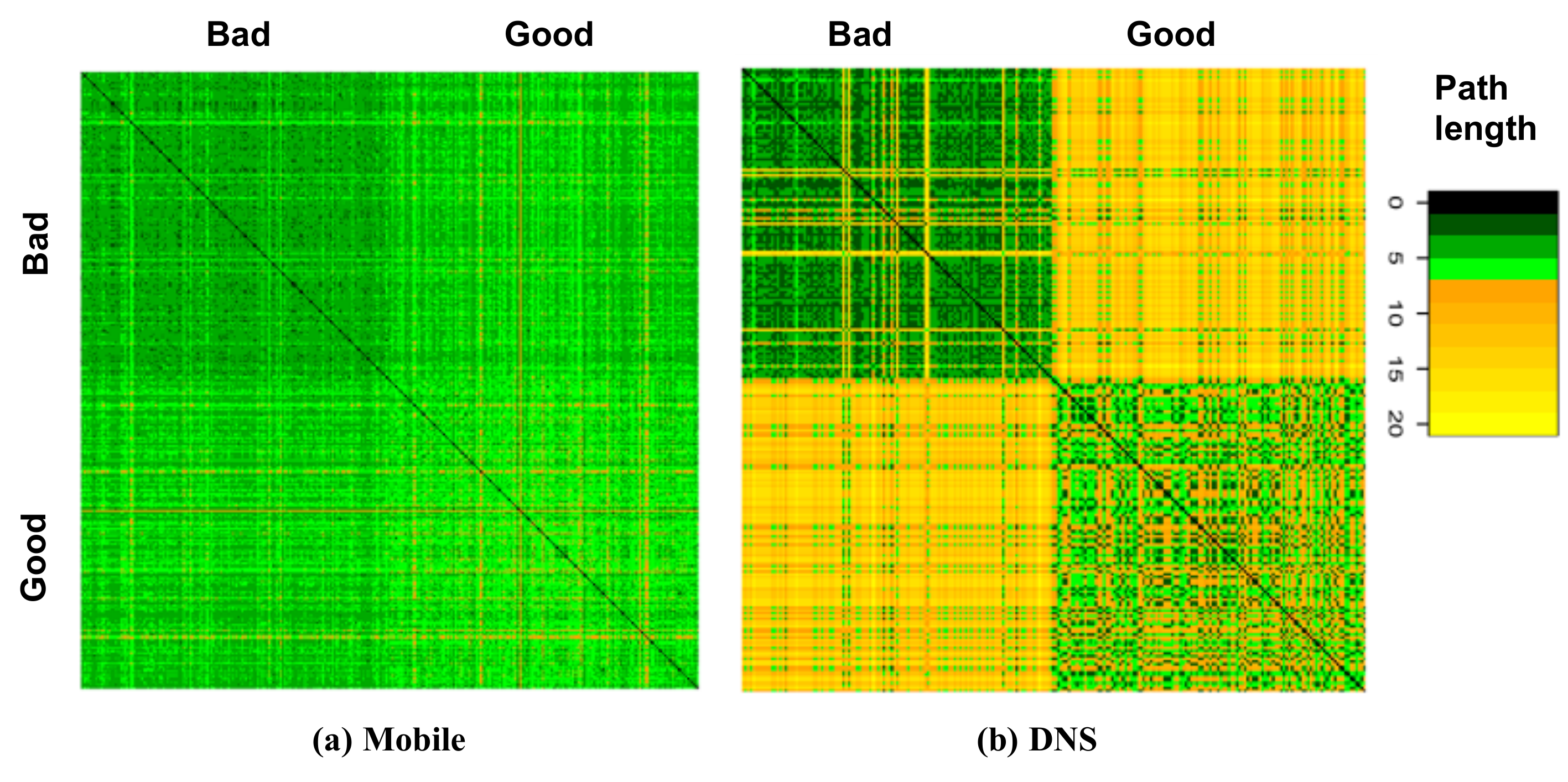,width=0.38\textwidth}%\vspace{-.15in}
\caption{Shortest path length matrix between bad/good and bad/good nodes}\label{heatmap}}

\end{center}
\end{figure}

Fig.\ref{heatmap} provides the matrix representing the shortest path lengths between nodes in $C_B$ and $C_G$. The range of lengths are from 0 to 20, each of which is illustrated as a color between black (0) and yellow (20) in the matrix. The figure shows a few important observations. First, generally in both datasets,  intra-cluster distances are smaller (darker and greener colors in the figure) than inter-cluster distances between $C_B$ and $C_G$. Second, $C_B$'s intra-cluster distances are the lowest (i.e., the darker and greener color in the figure) in both datasets. Finally, the difference between intra-cluster and inter-cluster distances in $DNS$ is much larger than that in $Mobile$.

\begin{figure*}[htbp]
\begin{center}
\parbox{1.0\textwidth}{
\centering 
\subfigure[CDF between good nodes]{
\psfig{file=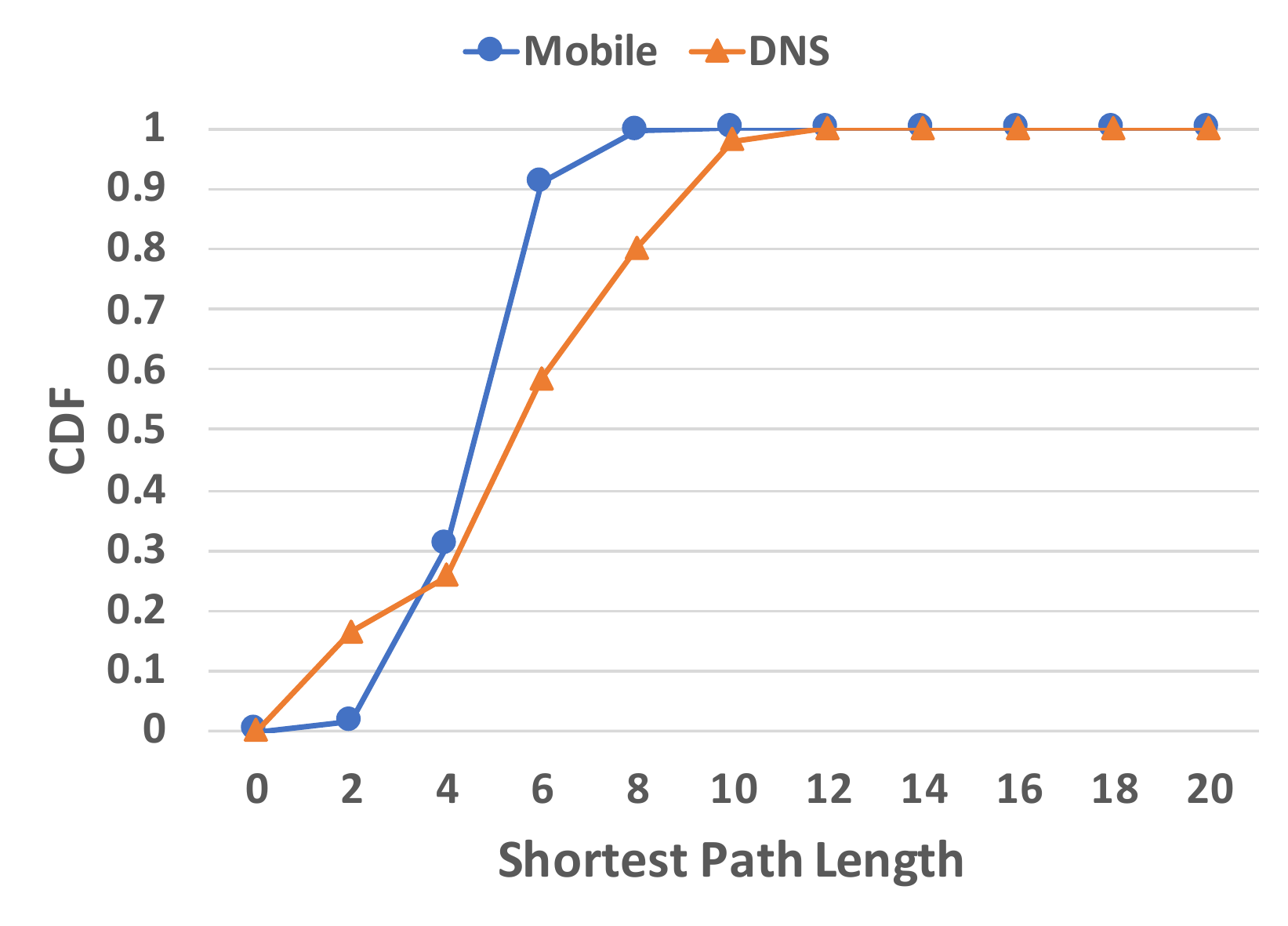,width=0.3\textwidth, height=0.18\textheight}\label{benign_cdf}}
\subfigure[CDF between bad nodes]{
\psfig{file=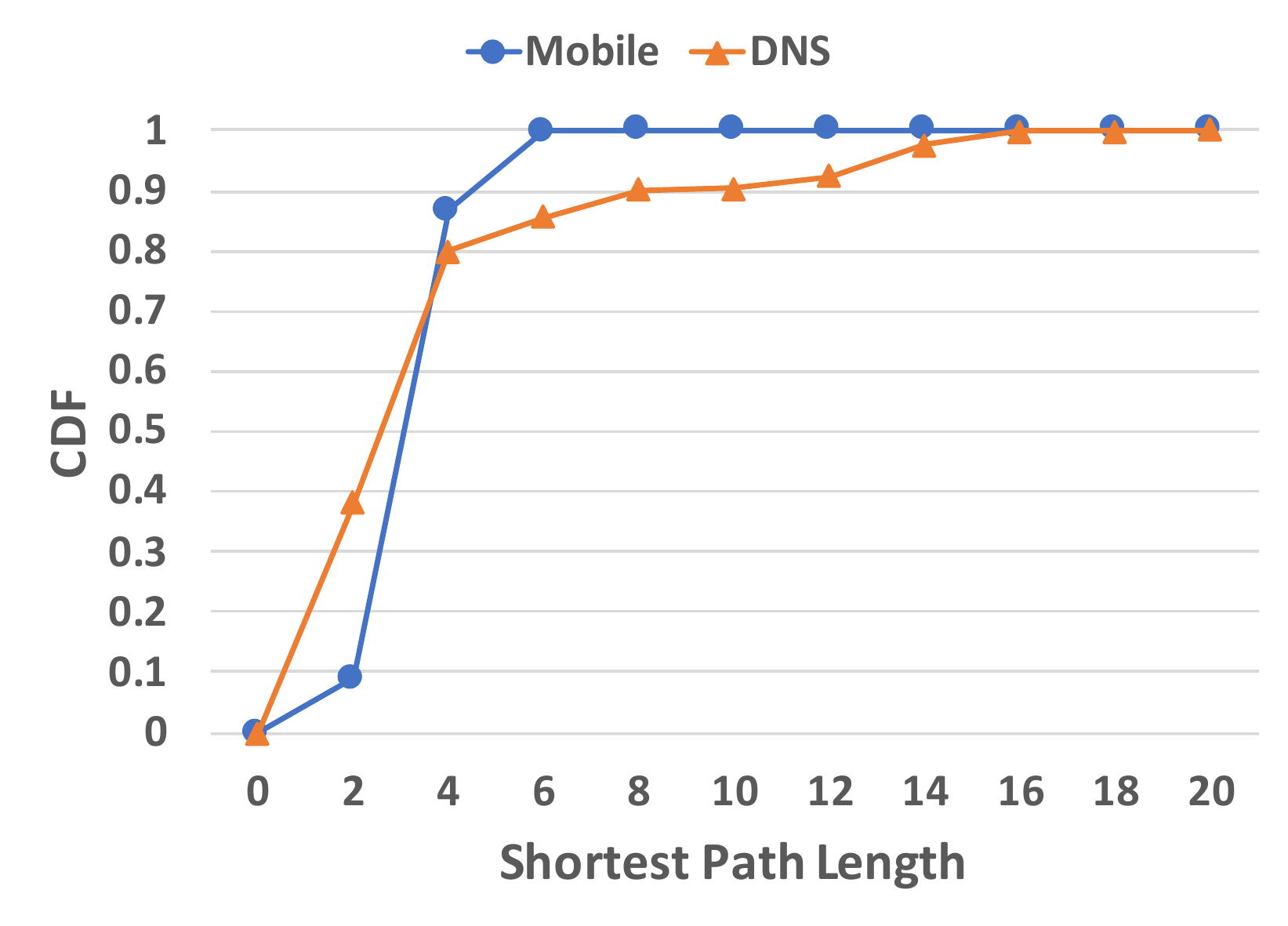,width=0.3\textwidth, height=0.18\textheight}\label{mal_cdf}}
\subfigure[CDF between bad and good nodes]{
\psfig{file=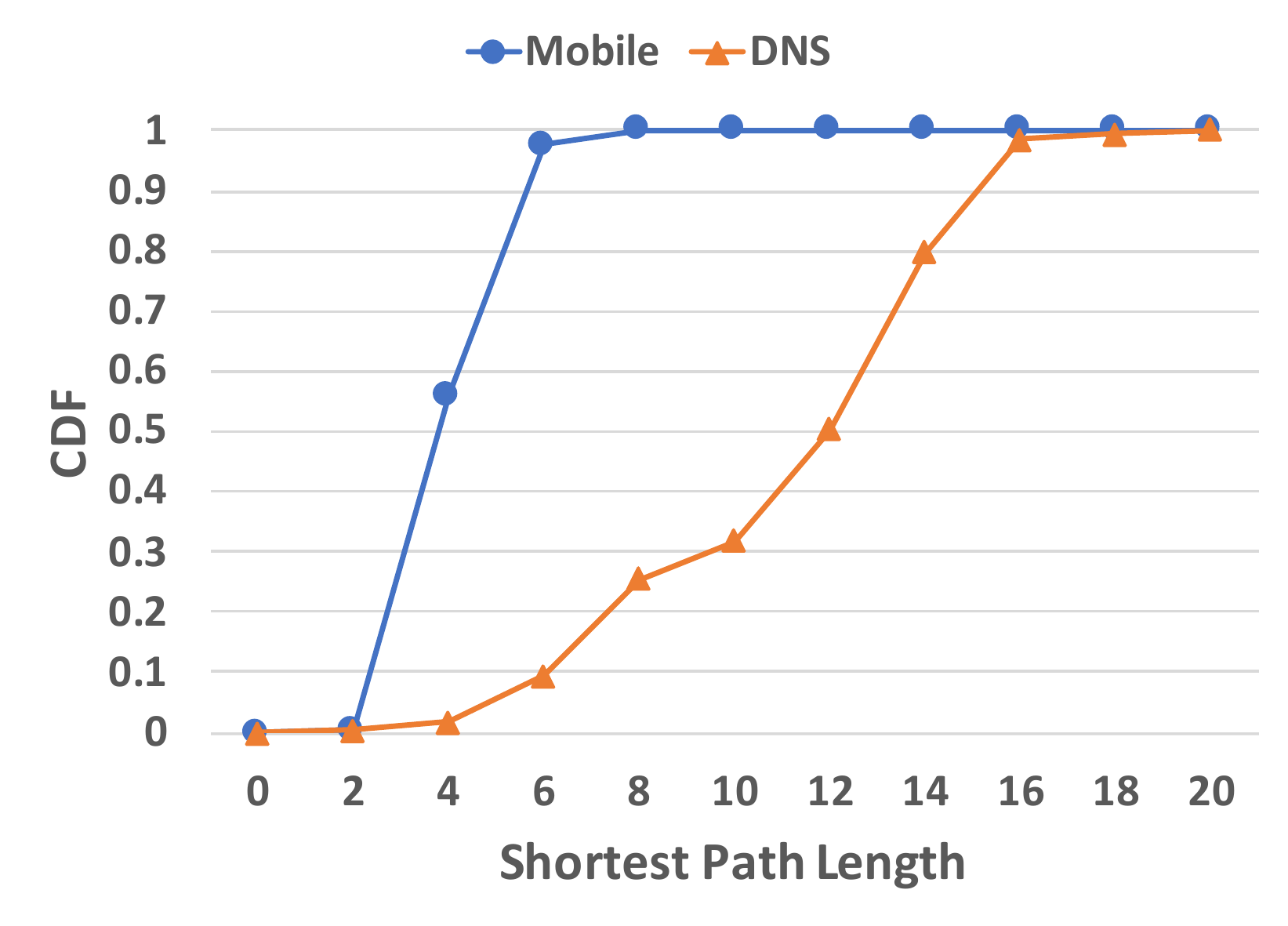,width=0.3\textwidth, height=0.18\textheight}\label{benign_mal_cdf}}
}%\vspace{-.17in}
\caption{Shortest path length CDF between bad/good and bad/good nodes}\label{short_path}

\end{center}
\end{figure*}
Fig.\ref{short_path} presents the CDF of shortest path lengths between nodes in $C_B$ and $C_G$, where the x-axis represents the shortest path lengths 
and the y-axis represents the corresponding CDF (i.e., portion of node pairs); one line per each dataset.  As shown in Fig.\ref{short_path}, the maximum lengths are 8 and 20 in Mobile and DNS, respectively.
Interestingly, $C_B$'s intra-cluster distances are similar in the two datasets. Specifically, 86.7\% of path lengths are within 4 (i.e., 2 or 4) in $Mobile$, and 79.9\% of path lengths are within 4 in $DNS$. 

On the other hand, we observe different characteristics in $C_G$'s intra-cluster distances, and inter-cluster distance between $C_B$ and $C_G$ for each dataset. In \emph{Mobile}, 91\% of path lengths between nodes in $C_G$ are smaller than or equal to 6 and only 9\% of path lengths are greater than 6; which are in fact similar to the inter-cluster distance between $C_B$ and $C_G$ where 97.5\% of path lengths are smaller than or equal to 6 and only 2.5\% of path lengths are greater than 6. In \emph{DNS}, 60\% of path lengths between nodes in $C_G$ are smaller than or equal to 6, while 90\% of path lengths between nodes in $C_B$ and $C_G$ are more than 6. In other words, although the intra-cluster distance is smaller than the inter-cluster distance in both datasets (i.e., the homophily relationships holds), the difference between intra-cluster and inter-cluster distances in 
\emph{Mobile} is relatively small. By contrast, the difference is relatively large in \emph{DNS}. On average, differences between $C_B$'s intra-cluster distance  and the inter-cluster distance  is only 0.9 in \emph{Mobile}, whereas the difference is 8 in \emph{DNS}, as shown in Table \ref{avg_short}.
\begin{table}[htb]%%
\centering
\small
%\begin{center}%
\begin{tabular}{|c|c|c|c|}\hline
\textbf{}& Good-Good  & Bad-Bad & Bad-Good\\
\hline
\textbf{Mobile} & 5.542 & 4.089 & 4.922\\
\hline
\textbf{DNS}  & 6.43 &4.523 &12.062\\
\hline
\end{tabular}

\caption{Average shortest path length}
\label{avg_short}
%\end{center}

\end{table}

Recall how the path length and  $\epsilon$ affect the behavior of BP. Relatively long inter-cluster distance (i.e., 12) diminishes the impact of bad (good) domains on good (bad) domains, irrespective of $\epsilon$ in \emph{DNS}. On the other hand, $\epsilon$ plays a big role in classification accuracy in \emph{Mobile}, due to the small differences between intra-cluster and inter-cluster distances. Concretely, bad devices have more  impact on good ones when we use higher $\epsilon$, resulting in the higher false positives. Hence, it is required to carefully choose $\epsilon$ close to 0.5 (e.g., 0.51) to avoid high false positives.

\textbf{Closeness centrality (CC)} of a node measures the average length of the shortest paths from the node to others \cite{trafficgraph}. Concretely, $CC$, of node $u$ is computed as:

\begin{displaymath}
CC_u = (N-1)\big/\sum_v l(v,u),
\end{displaymath}
where $N$ is the number of nodes in the graph and $l(v,u)$ is the shortest path length between $u$ and node $v$.

Essentially, CC takes into account both factors: the number of paths and the shortest path lengths. If all nodes in the graph are highly connected to each other with short path lengths, the CCs of all nodes will be similar. Indeed, the average CCs of bad and good devices in \emph{Mobile} are similar ($\approx$ 0.22) as shown in Table \ref{network_property}.  On the other hand, the average CC of bad domains is relatively small (0.088), compared to those of good and unknown domains (0.141 and 0.113, respectively) in \emph{DNS}. 
\begin{table}[htbp]%%

\begin{center}%
\begin{tabular}{|p{3cm}||p{1.7cm}|p{1.7cm}|}\hline
\textbf{}&\textbf{Closeness Centrality}  &\textbf{Eigenvector Centrality} \\
%Mobile-G.1 (Bad) & 0.253  & 0.035 \\
%\hline
%Mobile-G.1 (Good) & 0.25& 0.0258\\
%\hline
%Mobile-G.1 (Unknown) & 0.239& 0.015\\
\hline\hline
Mobile(Bad) & 0.223 & 0.0087\\
\hline
Mobile (Good) & 0.219& 0.0022\\
\hline
Mobile (Unknown) & 0.219& 0.0031\\
\hline\hline
DNS (Bad) & 0.088  & 0.005\\
\hline
DNS (Good) & 0.141 & 0.179 \\
\hline
DNS (Unknown) & 0.113  & 0.006 \\
\hline
\end{tabular}

\caption{Network properties of the Mobile and DNS datasets}
\label{network_property}
\end{center}

\end{table}
Along with the average shortest path given in Table \ref{avg_short}, we can conclude that the bad nodes in \emph{DNS} are much farther from other nodes and have less number of paths to other nodes, while good nodes are highly connected to good or unknown nodes, which is expected. This is because good domains are not likely to have many connections to bad domains, but have many connections to good or unknown domains. Hence, the classification accuracy is not sensitive to $\epsilon$ in in \emph{DNS}.

\textbf{Eigenvector centrality (EC)} of a node measures its influence in the graph. Concretely, $EC$ of node $u$ is computed:

\begin{displaymath}
EC_u = \kappa_1^{-1}\sum_vA_{uv}EC_v,
\end{displaymath}
where $v$ is $u$'s neighbor, $A$ is the adjacency matrix of the graph,  $\kappa_1$ is its largest eigenvalue.

\begin{figure*}[ht]

\begin{center}
\parbox{0.9\textwidth}{
\subfigure[Mobile eigenvector centrality distribution]{
\psfig{file=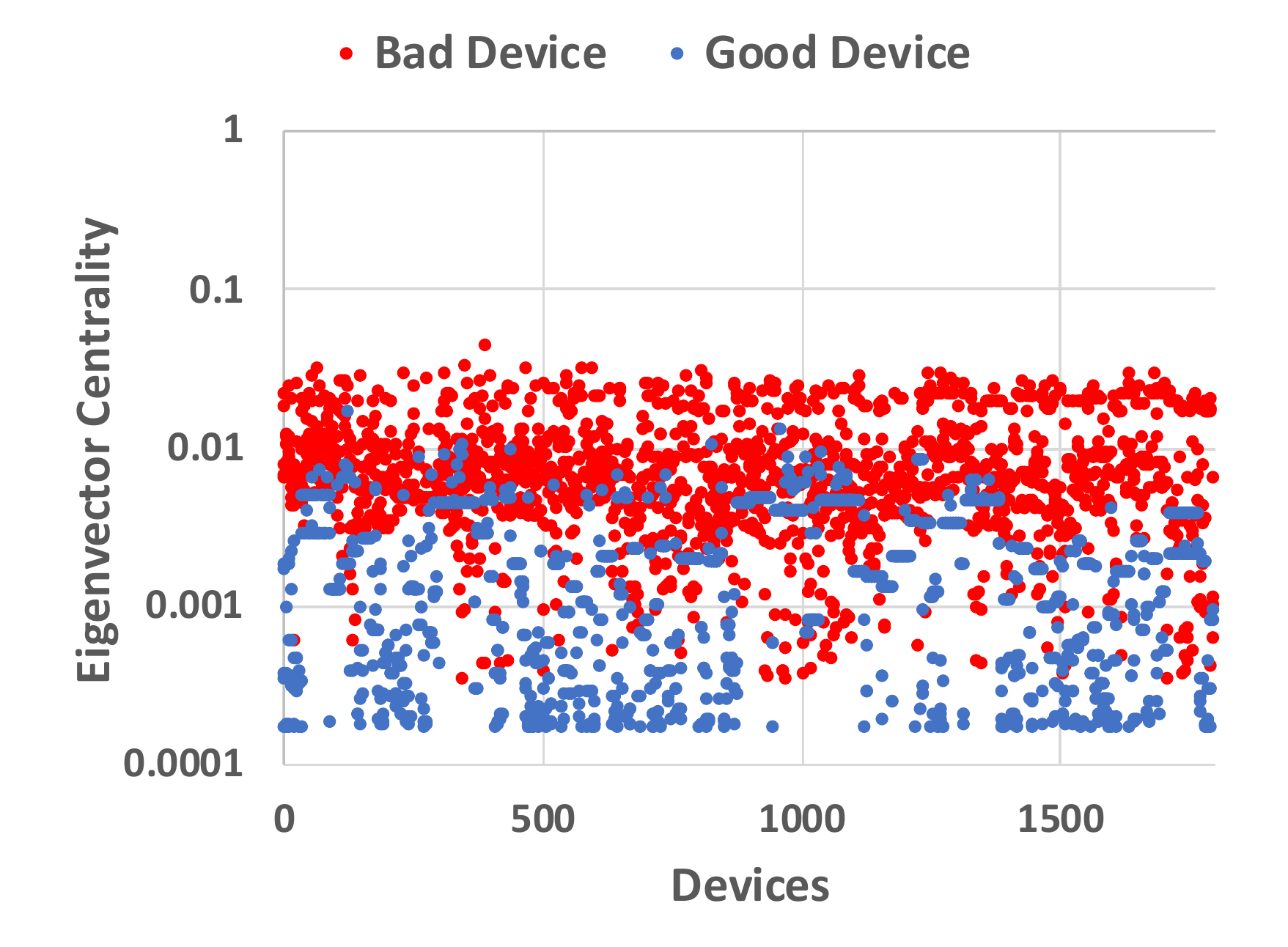,width=0.38\textwidth, height=0.2\textheight}\label{newbenign_eigen}}
\subfigure[DNS eigenvector centrality distribution]{
\psfig{file=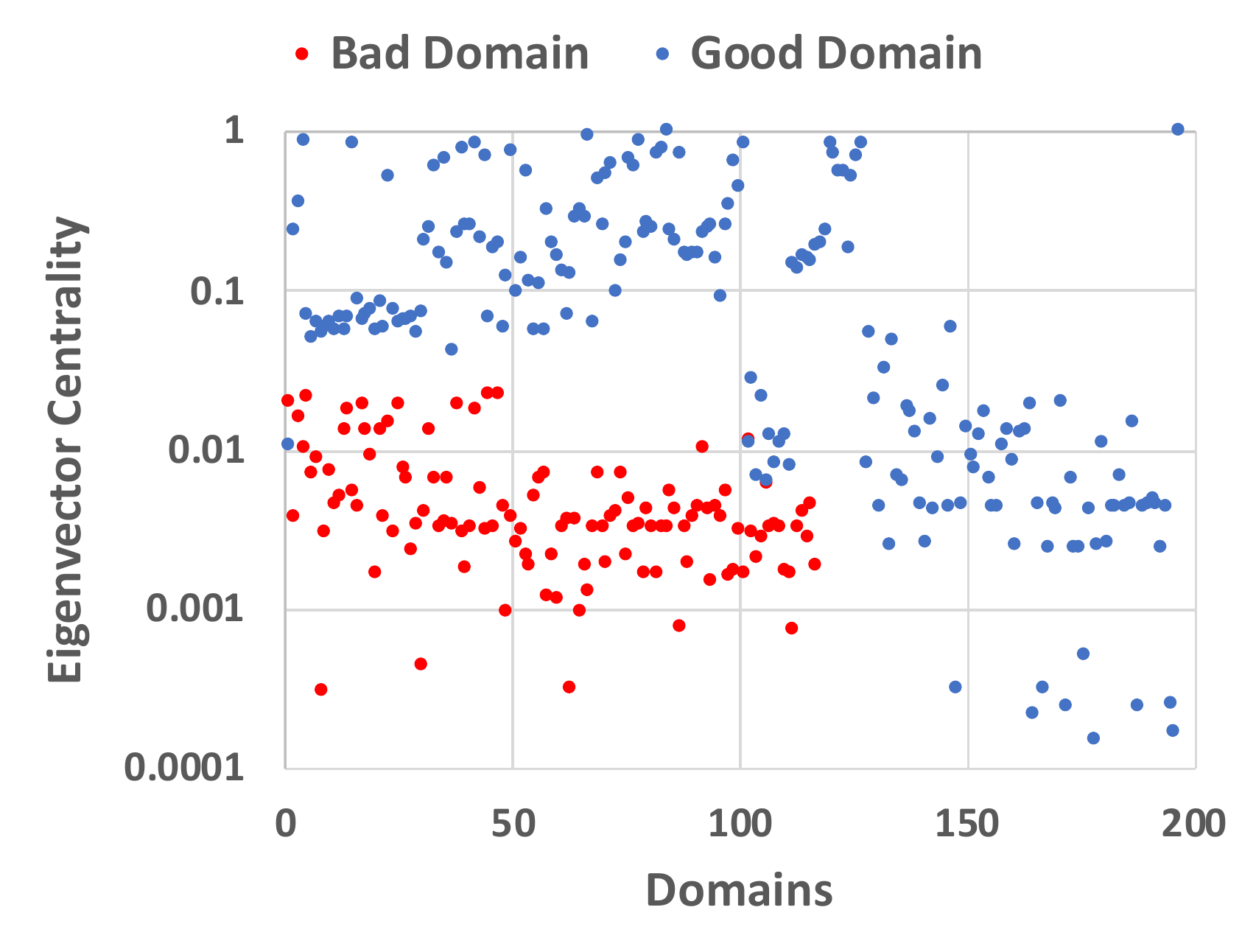,width=0.38\textwidth, height=0.2\textheight}\label{mada_eigen}}
}
\caption{Eigenvector centrality distributions of the Mobile and DNS datasets}\label{eigen}
\end{center}

\end{figure*}
A node with high EC means that it is highly connected to other \emph{influential} nodes.  That is, messages are most frequently passing through a node with high EC so that it will play a key role during belief propagation process. As shown in Table \ref{network_property}, there is clear difference on ECs between \emph{Mobile} and \emph{DNS} graphs. In general, the average ECs of bad, good, and unknown devices are almost similar (i.e., 0.0087, 0.0022, 0.0031, respectively) in \emph{Mobile} graph. This means that all nodes in the graph are highly connected with each other so that there are no significantly influential nodes in the graph. Note that the ECs of bad devices is the highest, meaning that as the higher $\epsilon$ is used, the score of bad devices can dominate the network, resulting in high false positives.
On the other hand, the average EC of good domains (0.179) are much higher than those of bad and unknown domains (0.005 and 0.006, respectively) in \emph{DNS} graph. 

Fig.\ref{eigen} shows the distribution of eigenvector centralities of bad and good nodes in each dataset. Similar to results from $CC$, Fig.\ref{mada_eigen} shows that bad domains in \emph{DNS} are significantly further from other nodes and are not connected to influential nodes, meaning that there is a smaller number of paths to other nodes. Although the $EC$s of good domains are high on average, they are well-distributed. This is in fact expected, as there can be influential and non-influential domains.  

By our definition in \emph{Mobile}, bad devices can have edges with all types of apps (i.e., bad, good, suspicious, and no-info apps); good devices  can have edges with good and no-info apps. This means that good devices could have a similar number of paths with both good and bad devices; bad devices, however,  have more paths with other bad devices  than good devices. Consequently, bad devices become relatively influential and connected to other influential bad devices, resulting in the relatively high $EC$s as shown in Fig.\ref{newbenign_eigen}. 

%On the other hand, good and unknown devices' $EC$s along with the results in Table  \ref{avg_short} suggest that good devices are not much different from unknown devices in terms of their topological locations. More specifically, good devices are

%However, good nodes are relatively farther from bad nodes in  \emph{Mobile-G.2}, compared to  \emph{Mobile-G.1}, as shown in Table \ref{avg_short}. This means that good nodes are less likely to connect to influential bad nodes, resulting in relatively lower $EC$s, as shown in Fig.\ref{newbenign_eigen}.

Recall how the number of paths and  $\epsilon$ affect the behavior of BP. One important observation in Fig.\ref{eigen} is that bad devices are more influential on others than good devices in \emph{Mobile}; whereas bad domains are less influential on others in \emph{DNS}. Along with results in Table \ref{avg_short}, we can conclude that bad devices get more influences from bad devices, especially from those influential bad devices, than good devices; so that good devices' messages have relatively less impact on bad devices. Consequently, there are not much change on false negatives, irrespective of $\epsilon$, as opposed to false positives.

\section{Post Analysis of Classified Devices}~\label{sec_post_analysis}
Recall that our goal is to identify unknown compromised devices whose owners often inadvertently install apps without much consideration of consequences. To further verify the accuracy of the classification results in Section ~\ref{accuracy}, we measure the private information leakage on classified devices (Section ~\ref{sec_leakage_analysis}) and study underlying network infrastructure accessed by the devices (Section ~\ref{sec_infratructure}).

\subsection{Privacy Leakage}\label{sec_leakage_analysis}
 It is known that devices having bad apps often leak private information~\cite{droidjust,credroid}. We examine samples of classified good and bad devices to study private information leakage on them. 

\textbf{Ethics}: It is important to note that our research is conducted on an anonymized version of the dataset where possible privacy concerns are carefully considered and addressed. Before analysis, all identifiable and personal information (e.g., phone number, user names) or device identifiers appearing in the traffic is anonymized or replaced with pseudo information.

We have compiled a list of private information which often leaks in mobile networks based on the previous research ~\cite{taintdroid,recon,mosaic,bugfix}. It has been shown privacy leaks often occur in a structured format, i.e., a key-value pair in HTTP headers ~\cite{recon,bugfix,locationtracking}. For example, a login password leaks with \emph{pwd=mypaSS123} where \emph{pwd} is a key and \emph{mypaSS123} is a value. Note that the key for a specific type of private information might be different depending on each app or device ~\cite{recon,bugfix}. From the dataset, we heuristically extract highly-related keywords to each type of private information. Examples of such keywords are summarized in Table.~\ref{pii_keywords} in the appendix. Note that we only present a few examples of keywords in Table.~\ref{pii_keywords}, as the keywords are similar yet small variations such as imei1 and imei7.
Note that we also validate that the keywords are in fact used to leak the corresponding type of information by running apps on the two mobile devices mentioned in Section~\ref{sec_data}. The private information might have been obfuscated in the traffic (e.g., hashing). However, it is shown that most information leakage occurs in plaintext in mobile networks~\cite{recon}. We thus argue that the following results represent the general behavior of each device. 

We sort unknown devices by their final beliefs, and choose top-100 devices with high scores as bad devices and bottom-100 devices with low scores as good devices. Note that, we choose scores derived from the results with parameters $vt=5$ and $\epsilon = 0.51$, based on discussion in Sections ~\ref{accuracy} and ~\ref{sec_graphstruct}. Then, we inspect HTTP packets originated from each device. Specifically, we search the keywords in each device's all of the HTTP headers, and consider a device leaks its private information if a non-empty key-value string corresponding to given keywords is found in any headers. 
\begin{figure*}[htbp]
\begin{center}
\parbox{0.9\textwidth}{
\centering 
\subfigure[Statistics for Private information leakage]{
\psfig{file=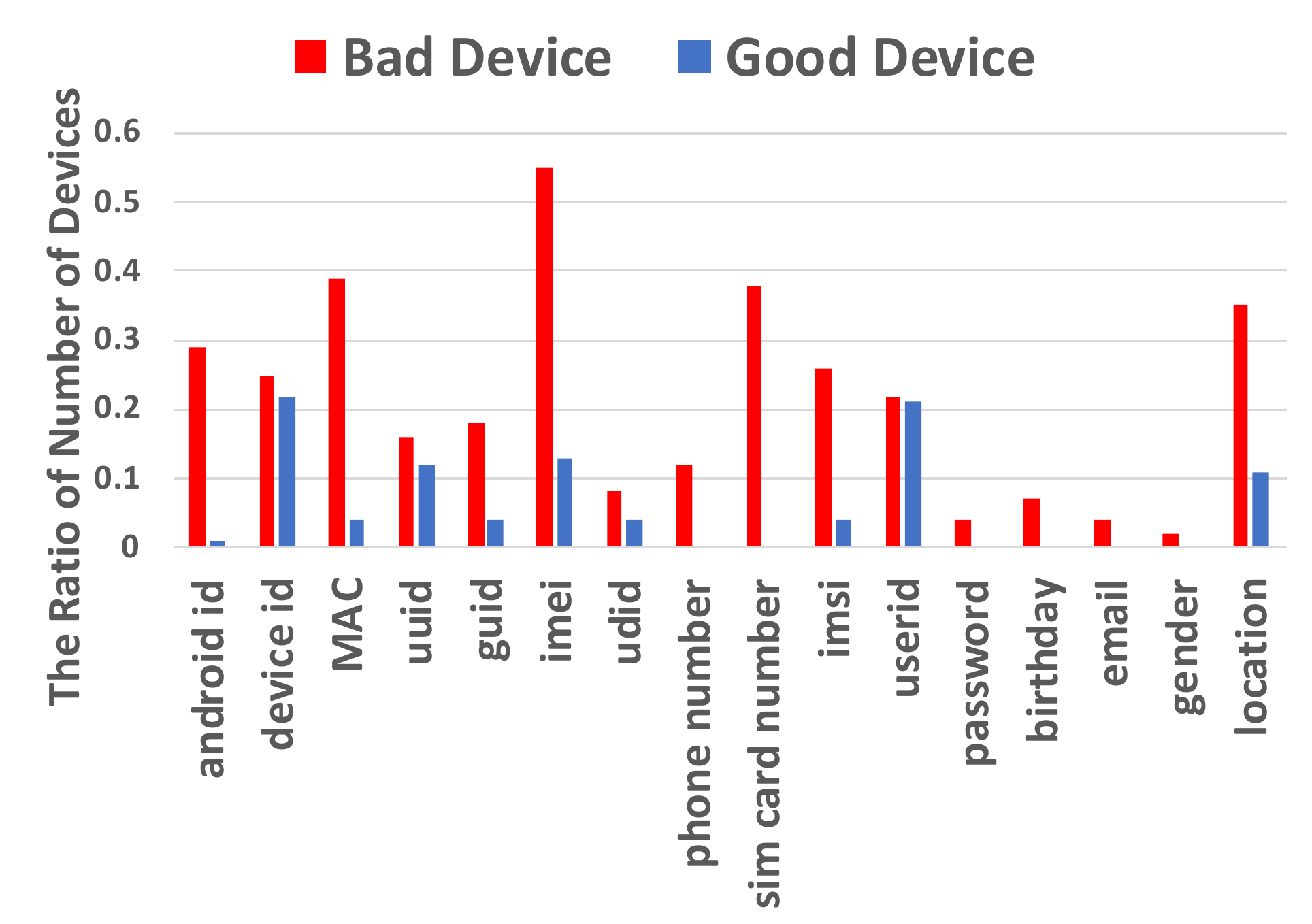,width=0.38\textwidth, height=0.2\textheight}\label{pii_detail}}
\subfigure[The CDF of the number of leaked information types]{
\psfig{file=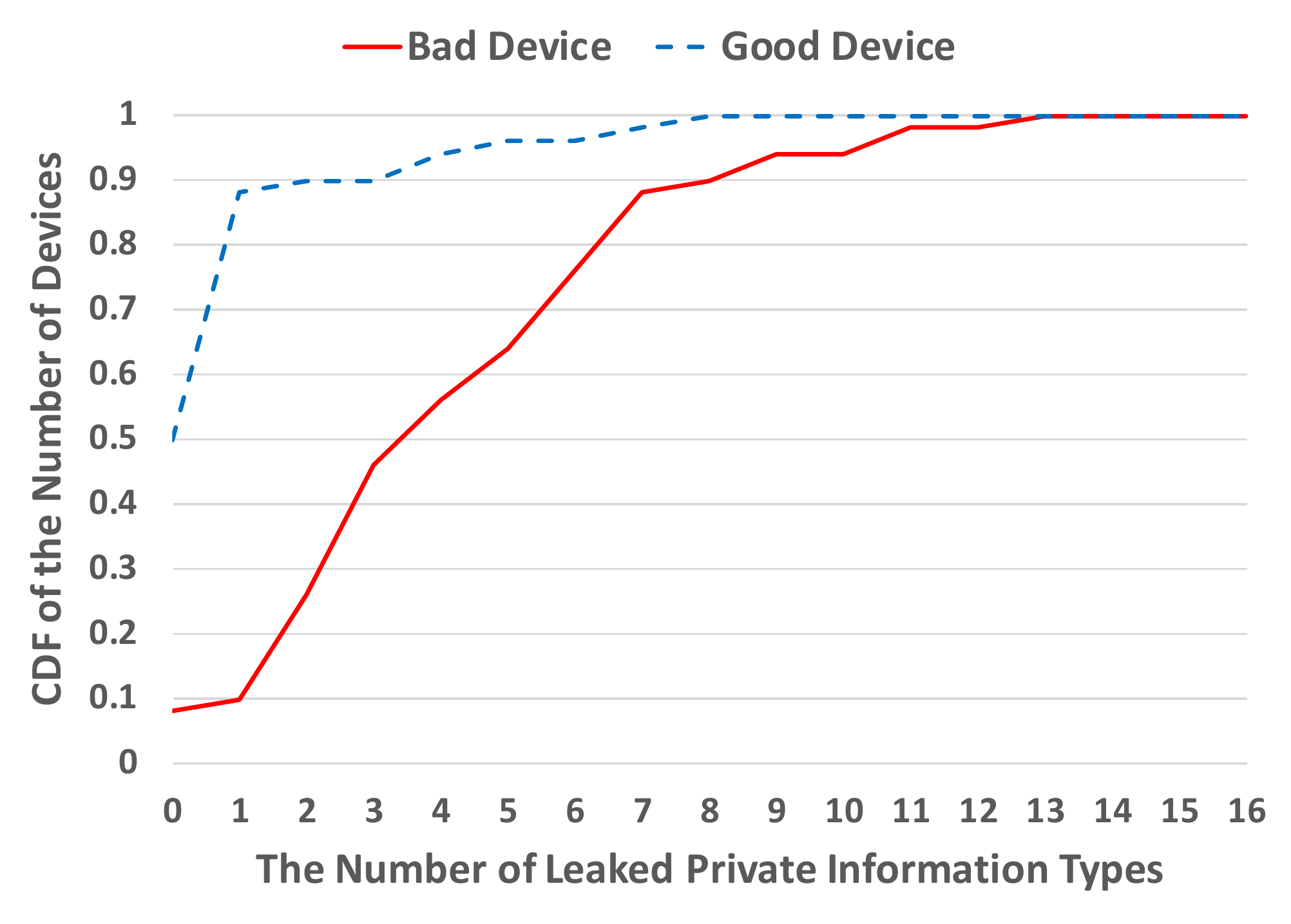,width=0.38\textwidth, height=0.2\textheight}\label{pii_cdf}}
}%\vspace{-.2in}
\caption{Private Information Leakage}\label{pii_app_traffic}
\end{center}

\end{figure*}

Fig.\ref{pii_detail} presents the ratio of the number of devices leaking each type of private information, where the x-axis represents each information type and the y-axis represents the ratio of the number of devices (i.e., the number of devices leaking corresponding information over the total number of devices). Generally, a large number of bad devices leak private information compared to good devices. Notably, only bad devices leak highly sensitive information including passwords and email addresses. Interestingly, some of the good devices also leak private information. Note that, as we select good devices with bottom-100 scores, it is less likely that these classified good devices include false negatives. 
%most of these devices do not belong to false negatives in classification results, as we choose bottom-50 devices. 
One possible reason is that even non-bad apps, particularly location-based searching apps, sometimes leak private information such as location data and device identifiers in order to support their functionality~\cite{taintdroid,recon}. In fact, we observe that the majority of leaking apps on good devices are location-based searching apps.

To compare leaking on good and bad devices, we measure (i) the number of leaked private information types, (ii) the distribution of the number of leaking apps and packets. Fig.\ref{pii_cdf} shows the CDF of the number of leaked private information type of devices, where the x-axis represents the number of leaked private information and the y-axis represents the CDF of the number of devices (i.e., corresponding portion of devices). As shown in the figure, half of good devices do not leak any information and 38\% of good devices leak only one information type. On the contrary, 92\% of bad devices leak at least one private information type and 36\% of bad devices leak more than 5 information types.

\begin{figure*}[htbp]
\begin{center}
\parbox{1.0\textwidth}{
\centering 
\subfigure[CDF of the leaking app ratio]{
\psfig{file=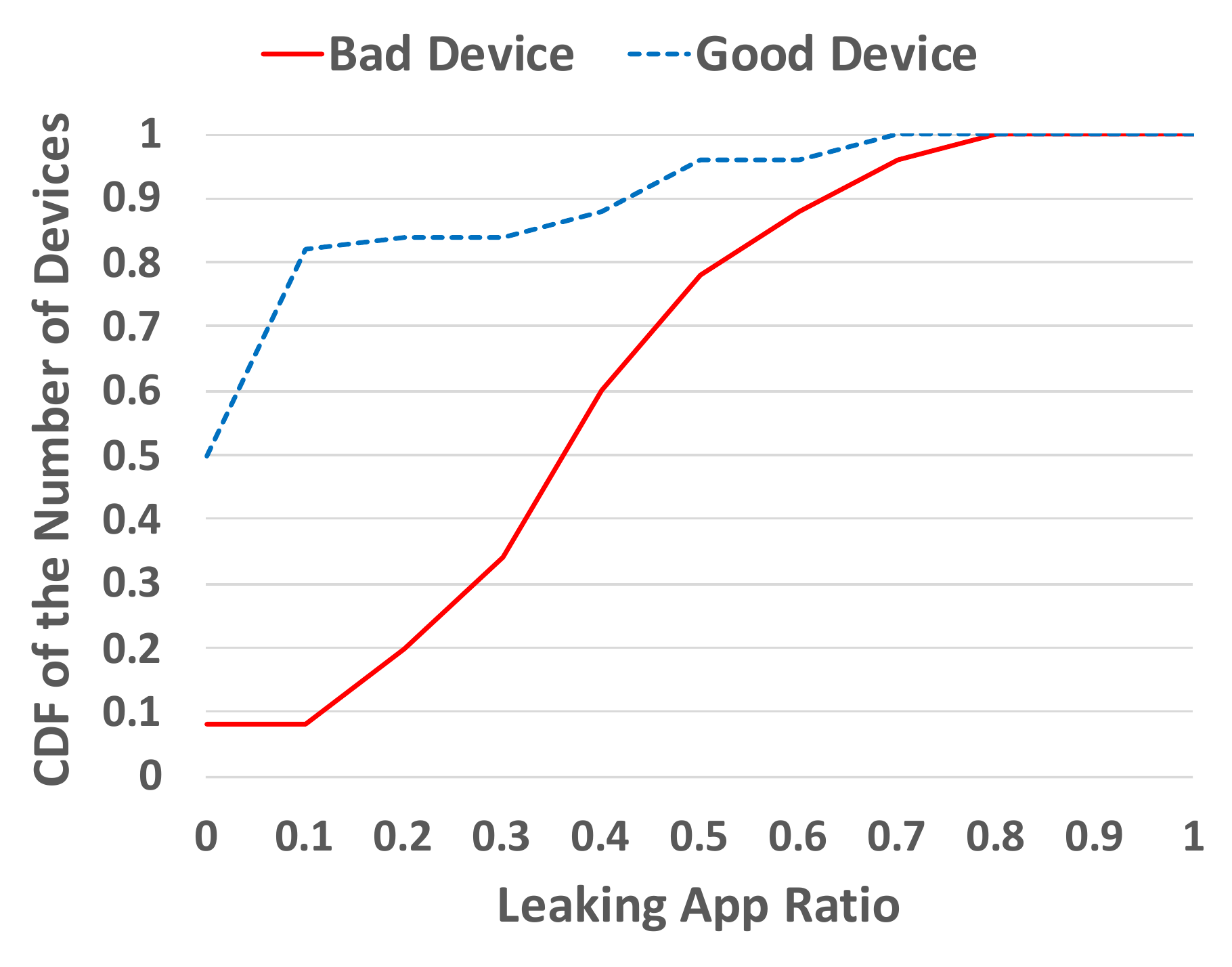,width=0.23\textwidth, height=0.2\textheight}\label{pii_app_cdf}}
\subfigure[The number of leaking apps]{
\psfig{file=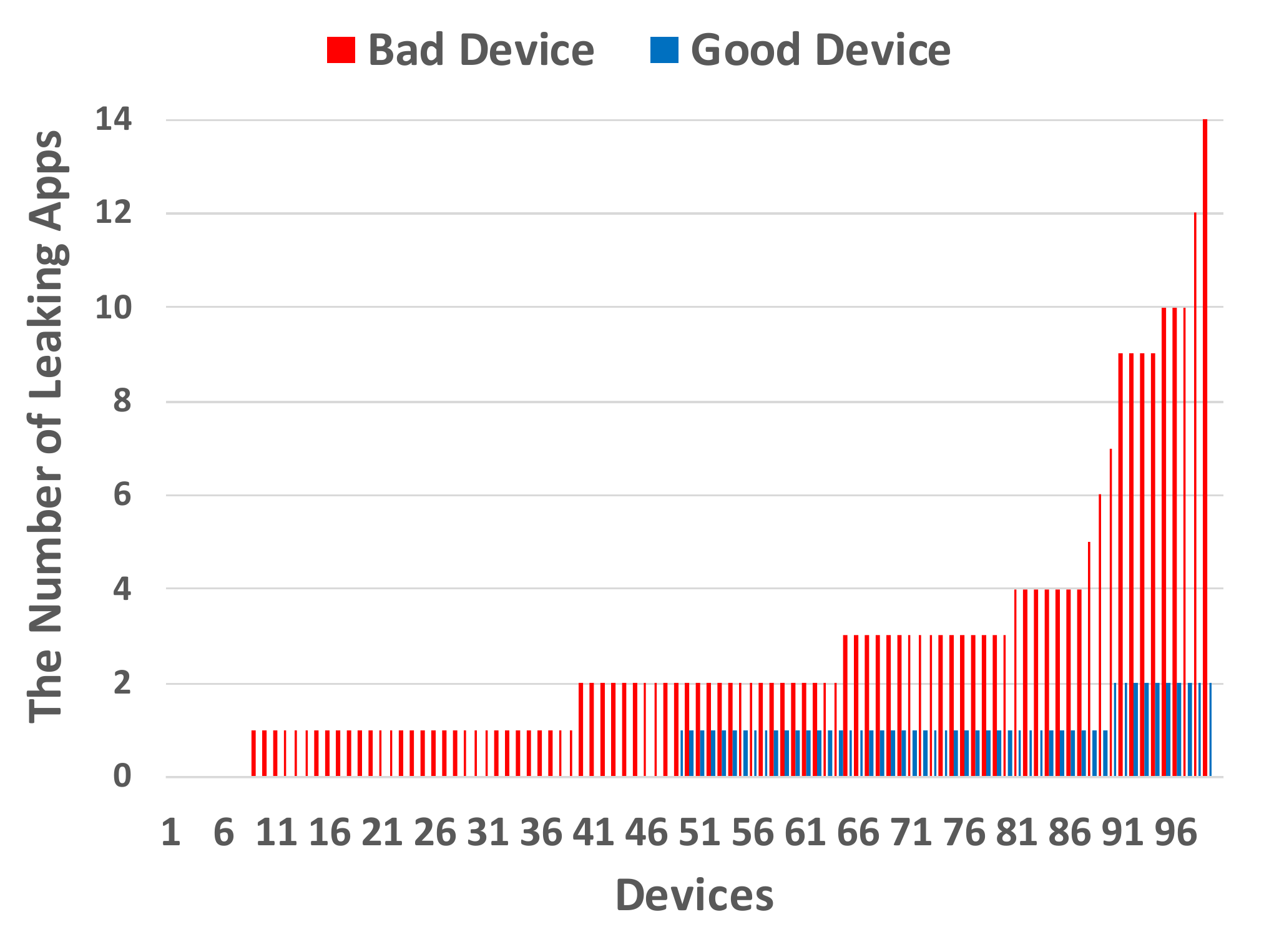,width=0.23\textwidth, height=0.2\textheight}\label{pii_app_cnt}}
\subfigure[CDF of the leaking packet ratio]{
\psfig{file=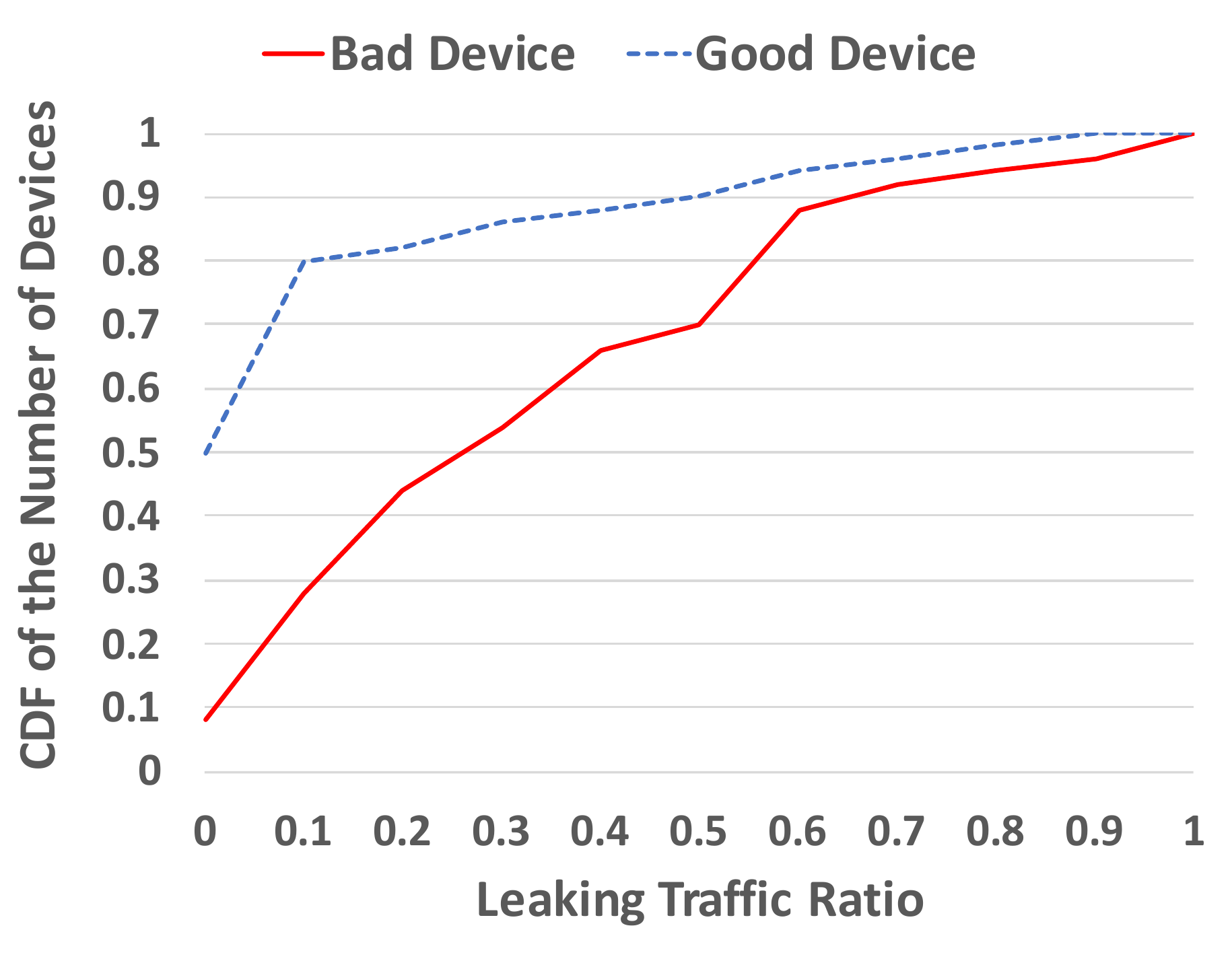,width=0.23\textwidth, height=0.2\textheight}\label{pii_traffic_cdf}}
\subfigure[The of number of leaking packets]{
\psfig{file=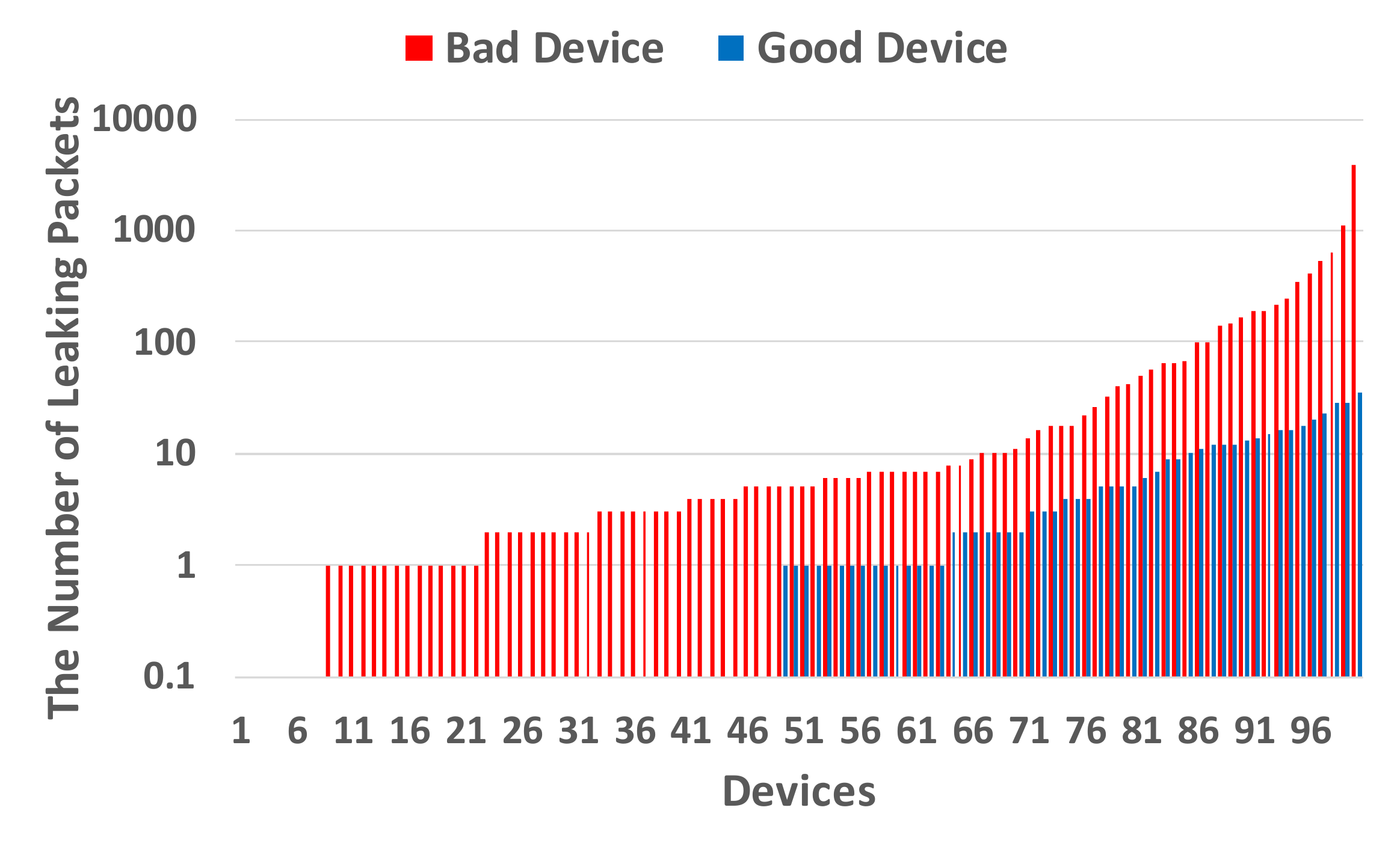,width=0.23\textwidth, height=0.2\textheight}\label{pii_traffic_cnt}}
}%\vspace{-.2in}
\caption{The distribution of leaking apps and packets of devices}\label{pii_app_traffic_leak}
\end{center}

\end{figure*}

We further measure how many apps and packets of each device leak private information.  Fig.\ref{pii_app_cdf} and Fig.\ref{pii_traffic_cdf} present the CDFs of the leaking app  and the leaking traffic ratios, respectively. The x-axis in each figure represents the leaking app and leaking traffic ratios, respectively; the y-axis represents the portion of devices.

Leaking app ratio is measured by the number of apps leaking information over the total number of apps of each device; leaking traffic ratio is measured by the number of packets leaking information over the total number of packets of each device. As shown in the figures, although some good devices also leak private information, the ratios of leaking apps and packets of each device are relatively small. Specifically, we observe that among all the good devices that leak private information (i.e., 50\% of the good devices), 30\% of them have less than 10\% of leaking apps (Fig.\ref{pii_app_cdf}); the traffic of 41\% of them has less than 10\% of leaking packets (Fig.\ref{pii_traffic_cdf}).

Fig.\ref{pii_app_cnt} and Fig.\ref{pii_traffic_cnt} present the distribution of the number of leaking apps and packets of devices, respectively. The x-axis represents each device; the y-axis in each figure represents the corresponding number of leaking apps and packets of each device, respectively. Note that 50\% of good devices and 8\% of bad devices do not leak any information so that their numbers of leaking apps and packets are 0, which are not shown in the figures.

Interestingly, these leaking apps are not necessarily the same set as the bad apps in the ground truth set in Section~\ref{sec_data}. In fact, 85\% of these leaking apps are not the apps originally flagged as bad. In other words, the BP based inference relies on a largely independent set of apps compared to the ground truth to detect bad devices.  
%Nabeel: the above also clarifies one reviewer's claim on the baised inference.

Notably, we can see a clear difference between good and bad devices in terms of the number of leaking apps and packets. Specifically, if any, good devices have only one or two leaking apps; whereas 35\% of bad devices have more than two leaking apps. Also, although some good devices leak information, the number of leaking packets are less than 30; whereas 23\% of bad devices have more than 30 leaking packets.

%This result lends support to our claim that our approach in fact correctly detects good devices and bad devices. 
%Nabeel: since we mention this at the end of this subsection, we can omit this line here.
%captures the level of a device being compromised and the inadvertent behavior of its owner.%(the device)
%alley: I think this solves the doubt about overfitting problem. Because, the discovered devices are not necessarily having known malicious apps, but show bad behavior. We may need to explicitly put such contents here e.g., this result  also confirms that our classification does not suffer from overfitting problem..

 In summary, our privacy leakage analysis suggests that our approach can reliably detect unknown bad devices. Specifically, we show that although devices do not have bad apps from the ground truth, classified bad devices are showing undesirable behavior in terms of leaking their personal information. This result is also promising in that we can possibly identify unknown bad apps by our approach. Evidently, apps leaking sensitive information are not desirable and we may further analyze apps causing privacy leakage on the classified bad device. Since our focus is to identify devices, we leave the further investigation on apps as future work.

\subsection{Network Infrastructure Accessed}~\label{sec_infratructure}

\begin{figure*}[htbp]
\begin{center}
\parbox{0.9\textwidth}{
\subfigure[The CDF of the number of ASes utilized]{
\psfig{file=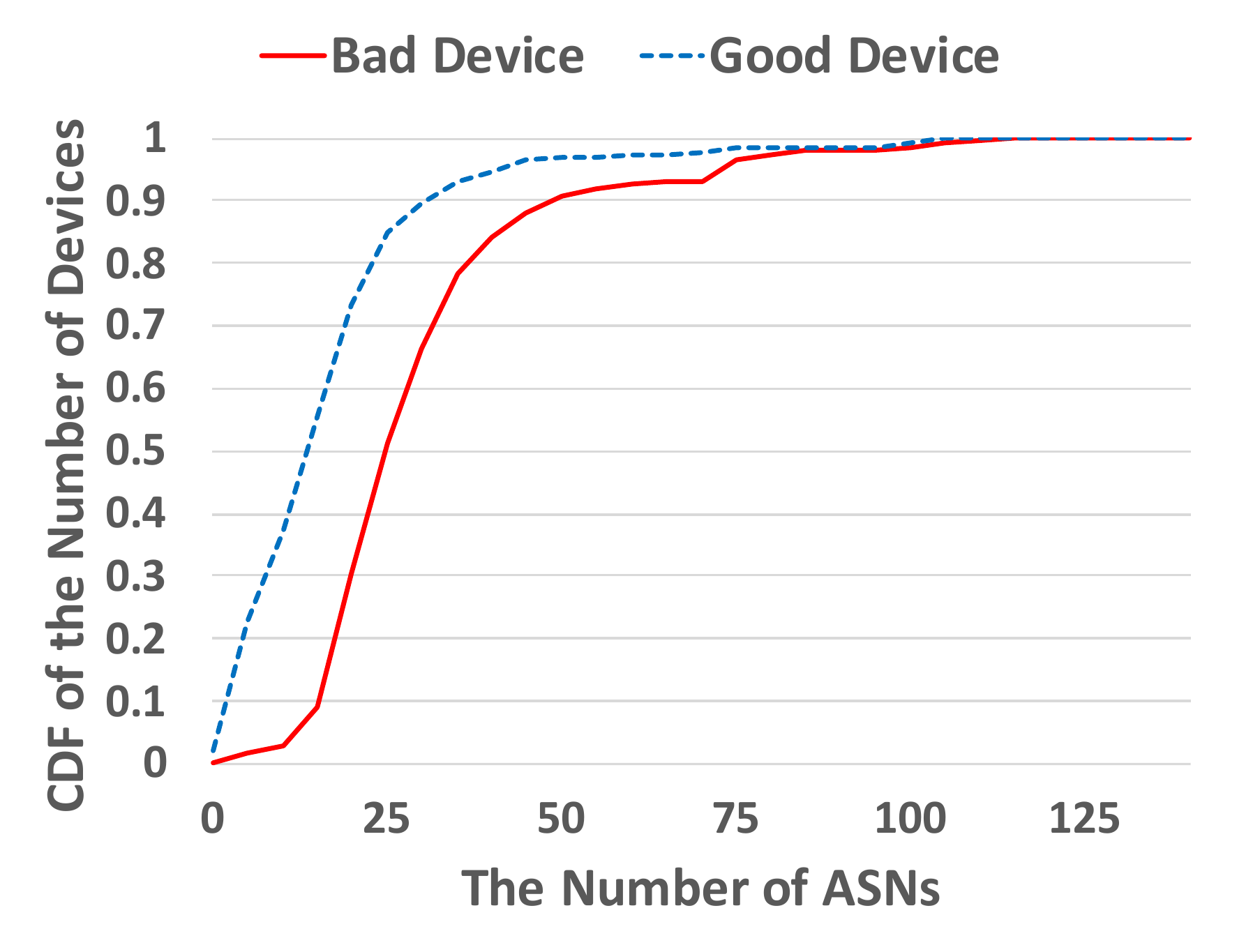,width=0.38\textwidth, height=0.18\textheight
}\label{fig:asn_cdf}}
\subfigure[The CDF of the number of short lived domains accessed]{
\psfig{file=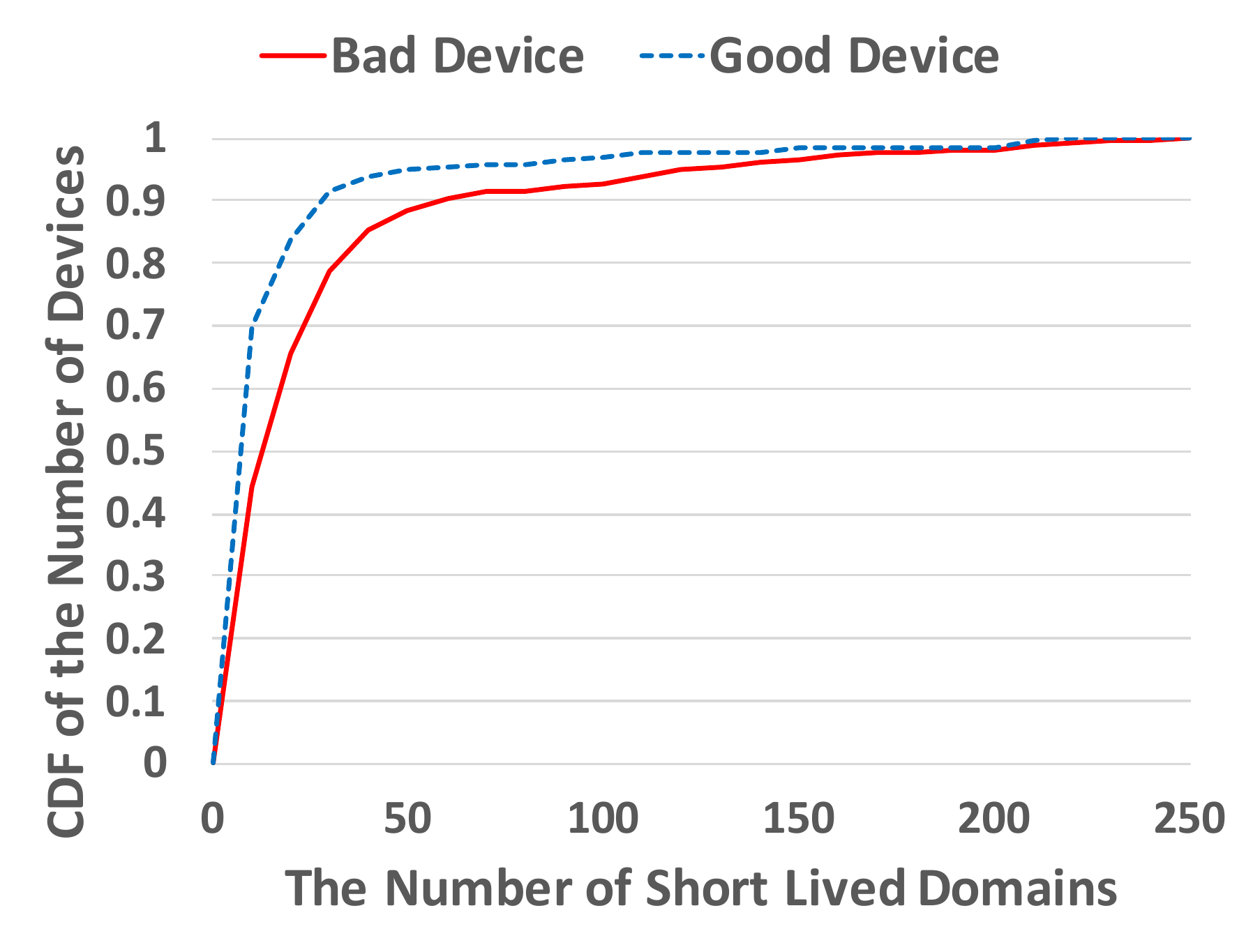,width=0.38\textwidth,height=0.18\textheight
}\label{fig:short_lived_cdf}}
}%\vspace{-.18in}
\caption{The CDFs of the number of ASes utilized and the number of short lived domains accessed}\label{network_infra}

\end{center}
\end{figure*}

We analyze the underlying network resources such as domains and IPs accessed by classified devices. It is well-known that miscreants utilize fast fluxing~\cite{fastflux:2008:NDSS}, where a given malicious domain is hosted at different IPs in a short period of time, to improve the availability of their malicious domains. Further, it is recently shown that miscreants move their domains from one hosting provider to another frequently to evade take down~\cite{codaspy_issa}. In our dataset, 94\% of hosting providers possess a single AS (Autonomous System). Thus, the above observation on hosting providers can be generalized to ASes. In this experiment, we seek to find out if indeed this AS behavior exists in the IPs accessed by the apps in the classified devices. Fig.~\ref{fig:asn_cdf} shows the CDF of the number of ASes utilized to host domains accessed by classified good and bad devices. Inline with above observations from previous research, it shows that bad devices tend to access IPs from more ASes compared to good devices. The figure shows, more than 90\% of bad devices access IPs from more than 20 ASes, whereas only 30\% of good devices exhibit the same behavior. 
As it is economical to create domains, nowadays, miscreants use many disposable domains to launch their attacks~\cite{reg:2017:RAID}. We thus also explore whether domains accessed by bad devices exhibit such a behavior. First, for the domains in the dataset, we extract the first seen and last seen dates of each domain from Farsight passive DNS repository~\cite{DNSDB}, which collects DNS queries resolved world-wide and serves historical DNS query data since 2011. We define a short lived domain as one whose DNS footprint is less than 3 months. Most of these domains are usually taken down, sink holed, or black listed, if identified malicious~\cite{takedown:2019:NDSS}. Fig.~\ref{fig:short_lived_cdf} shows the CDF of the number of short lived domains accessed by good and bad devices. The figure confirms with the previous research findings where, in general, classified bad devices access more short lived domains compared to good devices. It shows that 20\% of bad devices access more than 40 short lived domains, whereas only less than 10\% of good devices exhibit the similar behavior.

\section{Discussion and Limitations}\label{discussion}

\begin{figure}[tb]
\begin{center}
\parbox{0.42\textwidth}{
\centering
    \epsfig{file=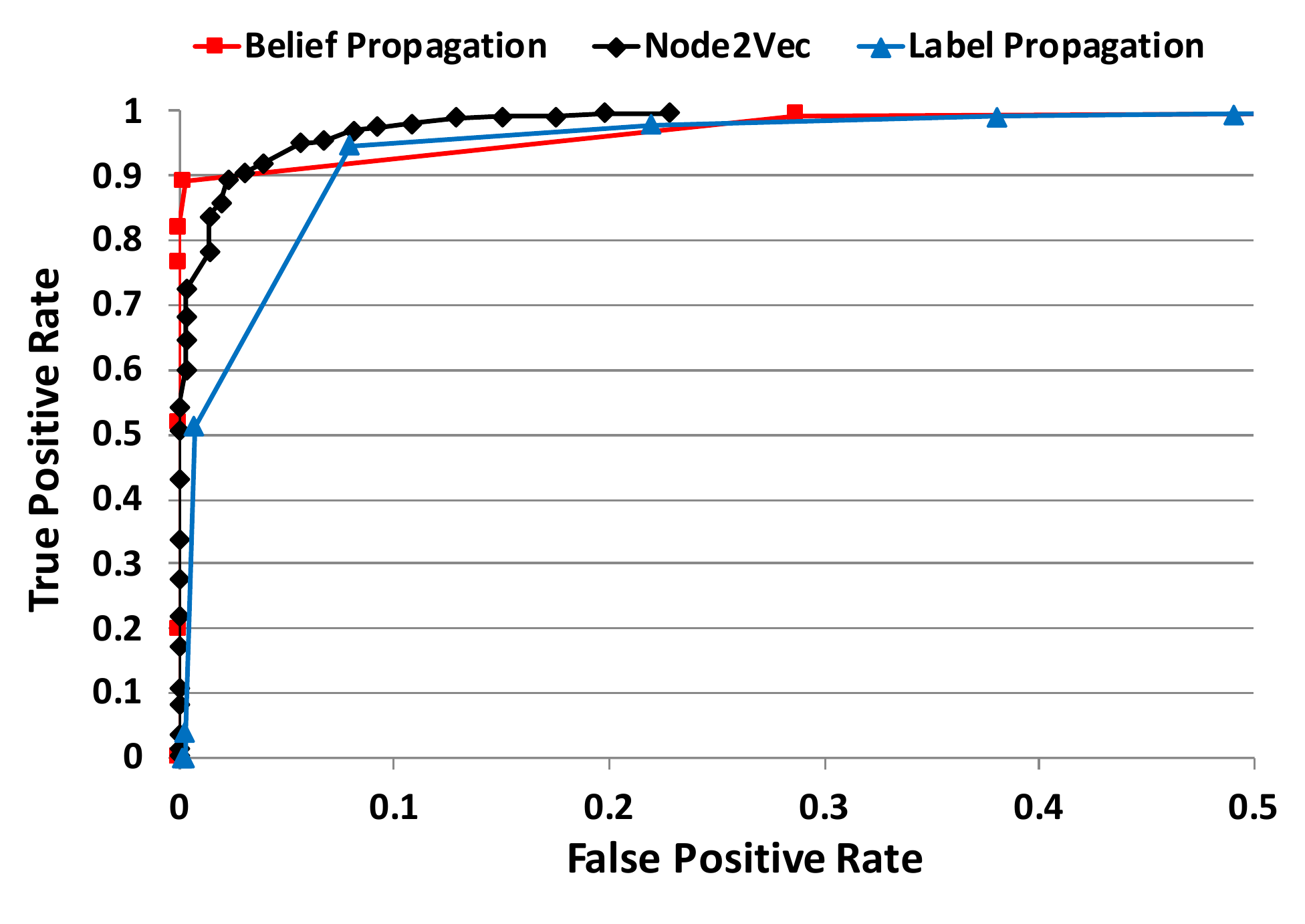,width=0.38\textwidth, height=0.18\textheight}%f\vspace{-.21in}
\caption{A ROC curve for three graph-inference algorithms}\label{compare_algo_app}}

\end{center}
\end{figure}

\heading{Baseline Approaches.} 
A naive baseline approach of utilizing a blacklist is restrictive and unable to detect compromised devices having previously unknown bad apps. In fact, our approach in general can  detect twice as many unknown bad devices not in the ground truth. An important consideration at early stages of detection is to identify which predictive model would work well to infer compromised devices from graph structured data. We evaluate three popular techniques: (1) Unsupervised node embedding along with Random Forest (Node2Vec)~\cite{node2vec}, (2) Label Propagation (LP)~\cite{lp:2002}, and (3) BP.  Fig.\ref{compare_algo_app} shows the ROC curves for the three approaches 
with the ground truth drawn with $vt = 5$. The ROC curves show that BP provides a low FPR, compared to LP and Node2Vec along with RF. There are a few possible reasons why BP performs slightly better: (a) node embedding based approach fails to capture labels into the embedding and may result in inaccurate classification when two or more nodes have similar structure but different labels, and (b) LP simply takes the average of the neighboring node values during each iteration and, unlike BP, it fails to capture the homophily relationships among neighboring nodes. Thus, it is not surprising that LP has the lowest performance out of the three approaches. Further, BP is several orders of magnitude faster than Node2vec.

\heading{App Strings and Filtering.}
Our study with the HTTP headers assumes that app strings revealed can characterize a specific app and its badness based on the prior research results ~\cite{beyondgoogleplay,usagepattern}. However, it is also known that malware writers often distribute repackaged apps by adding malicious codes to popular legitimate apps that may include the app string of such legitimate apps ~\cite{unknown_malice}. As we filter out popular apps to avoid false associations in BP, such repackaged apps may lead to false negatives; meaning that devices having only maliciously repackaged apps might be filtered and thus not detected. However, note that our approach does not use the knowledge of bad apps during the inference. In fact, we show that bad devices are detected independently from the known bad apps and our approach can be further extended to identify unknown bad apps by investigating the apps in detected devices in Section \ref{sec_post_analysis}. We thus argue that considering a limited number of known apps does not have a significant impact on the detection performance. 

\heading{HTTPS Encrypted Traffic.} Our approach can be applied both encrypted and not-encrypted traffic by extracting data in different ways. As mentioned in Section~\ref{sec_data}, however, a large percentage of mobile apps still rely on HTTP protocol for their communication for various reasons~\cite{inconsistent,networkframe,locationtracking,ample,lookat,bugfix}. It has also been observed in previous study that most malicious traffic is carried over HTTP for the same reasons~\cite{impending,nazca}. We thus believe that our approach with the high accuracy of detection (98\% AUC) using HTTP traffic can successfully identify the compromised devices in real-world mobile networks. 

Towards dealing with HTTPS, we also suggested a naive approach to use IP header only. While the detection accuracy is not promising as much as one to use HTTP, there is a line of work known as \emph{app fingerprinting} using IP headers or TCP traffic patterns that can be applied to improve the accuracy~\cite{realtime,appscanner,robust,tcp_finger}. That is, one may employ a supervised classifier to determine the specific app generating the observed HTTPS traffic~\cite{robust,appprint}. Once the apps in the traffic are identified, one can label the apps using external sources such as VT, and build a bipartite graph on which inference can be performed to detect compromised devices. We leave the investigation of this direction as future work.  

\heading{VT intelligence.} We used VT intelligence to label the apps to establish ground truth.  It has been pointed out in previous research that VT may have a limited coverage which could possibly bias results~\cite{unknown_malice,droppereffect}. We argue that the source of establishing ground truth is independent of our approach and an organization may get access to other sources~\cite{androzoo,koodous} which can possibly further reduce false positives and negatives. The goal of our approach is to start with a small set of ground truth based on any accessible intelligence source (VT or others) to expand the knowledge using graph inference to eventually identify devices which may have installed bad apps not detected previously. As discussed in Section~\ref{sec_post_analysis}, 85\% of leaking apps in detected devices are not the ones originally detected by VT, yet leaking a large amount of private information.

\section{Related Work}\label{sec_related}

\heading{Malicious App Detection.} Malicious app detection research falls into two categories:
code and network-analysis. Code analysis can further be categorized into static and dynamic analysis. Static analysis approaches derive signatures from app binaries based on features drawn from known malicious apps~\cite{droidminer,drebin}. However, it is often easy to evade detection by static analysis through code obfuscation and repackaging. Dynamic analysis monitors the behavior of an app such as privacy leakage or API calls~\cite{taintdroid,droidscope}. Unlike traditional desktop machines, however, it is often hard to perform run-time analysis on mobile devices because they are resource constrained. Network-analysis based approaches utilize traffic patterns such as packet sizes and detect anomalous network footprints ~\cite{deviation,trafficav,imbalanced,automated}. However, these approaches still extract each app-specific features which can easily be obfuscated. Also, they have the limitation that they cannot be applied for general purpose. Specifically, most proposed approaches are analyzing android-based apps, which cannot be directly used to detect iOS counterparts~\cite{taintdroid,robotdroid,androidmalware,droidminer}. By contrast, our approach identifies unknown compromised devices through graph inference without relying on device or app-specific features. Indeed, this is one of our key contribution that a network administrator does not need to do deep analysis on each individual device on the network, but can infer if an unknown device is compromised from other known devices. This also enables the administrator to quickly manage possible threats encountered at large-scale.

There are only a few research efforts that approach mobile security from a network administrator's point of view. Lever \emph{et al.} provided a large-scale network level analysis of mobile malware by investigating the DNS traffic of mobile devices~\cite{carrier}. This research is valuable in that authors show infection rate in real traces. However, authors did not provide a solution to identify mobile threats. Zhu \emph{et al.} proposed a method based on social network analysis to prevent worm propagation in cellular network~\cite{social}. The focus of this research was on worm propagation through MMS and SMS, which is different from our approach. Sharif \emph{et al.} propose a proactive approach that predicts if a user is connecting a malicious domain or web content by observing her mobile browsing behavior~\cite{impending}. In this paper, we focus on apps causing devices to be compromised rather than domains.
%focusing on compromised devices and malicious apps. 

%\subsection{Graph-inference Approaches}
\heading{Graph-inference Approaches.} A graph-inference approach has been employed in many different applications including anomaly detection, malware detection, fake social network account detection, fraud detection, and malicious domain detection. These applications construct different types of graphs including file-machine or file-relation graphs~\cite{polonium,marmite,guiltfile}, reviewer-product graphs~\cite{fraudeagle,speagle}, host-domain graphs~\cite{maldomain,enterprise_bp}, domain-IP graphs~\cite{codaspy_issa} and social network account graphs~\cite{Yuan:2019:fakeaccounts}. Although researchers have applied BP on a variety of applications, there is little study on the effect of BP parameters in different types of networks. In fact, most researchers either mentioned the results are not sensitive to BP parameters such as edge potentials~\cite{maldomain,codaspy_issa} or stated that specific values work well without any further description~\cite{polonium,guiltfile,fraudeagle}. 
However, we observe that the effectiveness of BP is relatively sensitive to characteristics of mobile networks. In Section~\ref{sec_graphstruct}, we thus discuss the unique characteristic of mobile networks in terms of their topology (i.e.,  where devices are closely connected to each other) compared to DNS based applications, and provide theoretical and experimental analysis on how such uniqueness may affect the results of BP.

\section{Conclusion}\label{sec_conclude}

We proposed a graph-inference based approach to identify compromised mobile devices. In doing so, we applied a well-known algorithm, BP, based on the intuition that devices sharing a similar set of apps will have a similar probability of being compromised. We studied this problem on real-world data that faithfully represents actual behavior of mobile users with which we demonstrate the effectiveness of our approach. 
We further study the impact of graph topology on BP parameters and highlight the distinct features of the mobile graph. Finally, our privacy leakage and hosting infrastructure post-analyses support the claim that our approach can reliably detect unknown compromised devices without relying on device-specific features. It is also important to take appropriate actions after compromised devices are detected. In fact, we also discuss that further investigation on detected devices might be helpful to identify unknown malicious apps, which we leave as a future work.

%\bibliographystyle{unsrt}  
%\bibliography{sigproc}
%\bibliography{references}  %%% Remove comment to use the external .bib file (using bibtex).
%%% and comment out the ``thebibliography'' section.

%%% Comment out this section when you \bibliography{references} is enabled.

\begin{thebibliography}{10}
\bibitem{australia-ISPs}
Graeme Burton.
\newblock Australia wants to force isps to protect customers from malware,
  2017.

\bibitem{australia2-ISPs}
James Eyers.
\newblock Cyber security minister says firms need to tell customers more about
  threats, 2017.

\bibitem{wired-ISPs}
Lily~Hay Newman.
\newblock Internet providers could be the key to securing all the iot devices
  already out there.

\bibitem{social}
Zhichao Zhu, Guohong Cao, Sencun Zhu, Supranamaya Ranjan, and Antonio Nucci.
\newblock A social network based patching scheme for worm containment in
  cellular networks.
\newblock In {\em Handbook of optimization in complex networks}, pages
  505--533. Springer, 2012.

\bibitem{carrier}
Charles Lever, Manos Antonakakis, Bradley Reaves, Patrick Traynor, and Wenke
  Lee.
\newblock The core of the matter: Analyzing malicious traffic in cellular
  carriers.
\newblock In {\em NDSS}, 2013.

\bibitem{cellbot}
Patrick Traynor, Michael Lin, Machigar Ongtang, Vikhyath Rao, Trent Jaeger,
  Patrick McDaniel, and Thomas La~Porta.
\newblock On cellular botnets: measuring the impact of malicious devices on a
  cellular network core.
\newblock In {\em Proceedings of the 16th ACM conference on Computer and
  communications security}, pages 223--234. ACM, 2009.

\bibitem{darkside}
Mauro Conti, Qian~Qian Li, Alberto Maragno, and Riccardo Spolaor.
\newblock The dark side (-channel) of mobile devices: A survey on network
  traffic analysis.
\newblock {\em IEEE Communications Surveys \& Tutorials}, 20(4):2658--2713,
  2018.

\bibitem{mobilesecindex2019}
Verizon.
\newblock Mobile security index.
\newblock 2019.

\bibitem{mosaic}
Ning Xia, Han~Hee Song, Yong Liao, Marios Iliofotou, Antonio Nucci, Zhi-Li
  Zhang, and Aleksandar Kuzmanovic.
\newblock Mosaic: Quantifying privacy leakage in mobile networks.
\newblock In {\em ACM SIGCOMM Computer Communication Review}, volume~43, pages
  279--290. ACM, 2013.

\bibitem{taintdroid}
William Enck, Peter Gilbert, Seungyeop Han, Vasant Tendulkar, Byung-Gon Chun,
  Landon~P Cox, Jaeyeon Jung, Patrick McDaniel, and Anmol~N Sheth.
\newblock Taintdroid: an information-flow tracking system for realtime privacy
  monitoring on smartphones.
\newblock {\em ACM Transactions on Computer Systems (TOCS)}, 32(2):5, 2014.

\bibitem{recon}
Jingjing Ren, Ashwin Rao, Martina Lindorfer, Arnaud Legout, and David Choffnes.
\newblock Recon: Revealing and controlling pii leaks in mobile network traffic.
\newblock In {\em Proceedings of the 14th Annual International Conference on
  Mobile Systems, Applications, and Services}, pages 361--374. ACM, 2016.

\bibitem{bugfix}
Jingjing Ren, Martina Lindorfer, Daniel~J Dubois, Ashwin Rao, David Choffnes,
  and Narseo Vallina-Rodriguez.
\newblock Bug fixes, improvements,... and privacy leaks.
\newblock 2018.

\bibitem{robotdroid}
Min Zhao, Tao Zhang, Fangbin Ge, and Zhijian Yuan.
\newblock Robotdroid: a lightweight malware detection framework on smartphones.
\newblock {\em Journal of Networks}, 7(4):715, 2012.

\bibitem{wild}
Adrienne~Porter Felt, Matthew Finifter, Erika Chin, Steve Hanna, and David
  Wagner.
\newblock A survey of mobile malware in the wild.
\newblock In {\em Proceedings of the 1st ACM workshop on Security and privacy
  in smartphones and mobile devices}, pages 3--14. ACM, 2011.

\bibitem{httpmalware}
Roberto Perdisci, Wenke Lee, and Nick Feamster.
\newblock Behavioral clustering of http-based malware and signature generation
  using malicious network traces.
\newblock In {\em NSDI}, volume~10, page~14, 2010.

\bibitem{deviation}
Asaf Shabtai, Lena Tenenboim-Chekina, Dudu Mimran, Lior Rokach, Bracha Shapira,
  and Yuval Elovici.
\newblock Mobile malware detection through analysis of deviations in
  application network behavior.
\newblock {\em Computers \& Security}, 43:1--18, 2014.

\bibitem{trafficav}
Shanshan Wang, Zhenxiang Chen, Lei Zhang, Qiben Yan, Bo~Yang, Lizhi Peng, and
  Zhongtian Jia.
\newblock Trafficav: An effective and explainable detection of mobile malware
  behavior using network traffic.
\newblock In {\em Quality of Service (IWQoS), 2016 IEEE/ACM 24th International
  Symposium on}, pages 1--6. IEEE, 2016.

\bibitem{machinelearning}
Fairuz~Amalina Narudin, Ali Feizollah, Nor~Badrul Anuar, and Abdullah Gani.
\newblock Evaluation of machine learning classifiers for mobile malware
  detection.
\newblock {\em Soft Computing}, 20(1):343--357, 2016.

\bibitem{imbalanced}
Zhenxiang Chen, Qiben Yan, Hongbo Han, Shanshan Wang, Lizhi Peng, Lin Wang, and
  Bo~Yang.
\newblock Machine learning based mobile malware detection using highly
  imbalanced network traffic.
\newblock {\em Information Sciences}, 433:346--364, 2018.

\bibitem{appscanner}
Vincent~F Taylor, Riccardo Spolaor, Mauro Conti, and Ivan Martinovic.
\newblock Appscanner: Automatic fingerprinting of smartphone apps from
  encrypted network traffic.
\newblock In {\em Security and Privacy (EuroS\&P), 2016 IEEE European Symposium
  on}, pages 439--454. IEEE, 2016.

\bibitem{drive}
Ruofan Jin and Bing Wang.
\newblock Malware detection for mobile devices using software-defined
  networking.
\newblock In {\em 2013 Second GENI Research and Educational Experiment
  Workshop}, pages 81--88. IEEE, 2013.

\bibitem{phishing}
Claudio Marforio, Ramya~Jayaram Masti, Claudio Soriente, Kari Kostiainen, and
  Srdjan Capkun.
\newblock Personalized security indicators to detect application phishing
  attacks in mobile platforms.
\newblock {\em arXiv preprint arXiv:1502.06824}, 2015.

\bibitem{beyondgoogleplay}
Haoyu Wang, Zhe Liu, Jingyue Liang, Narseo Vallina-Rodriguez, Yao Guo, Li~Li,
  Juan Tapiador, Jingcun Cao, and Guoai Xu.
\newblock {Beyond Google Play: A Large-Scale Comparative Study of Chinese
  Android App Markets}.
\newblock In {\em Proceedings of the Internet Measurement Conference 2018}, IMC
  '18, pages 293--307, New York, NY, USA, 2018. ACM.

\bibitem{maldomain}
Pratyusa~K Manadhata, Sandeep Yadav, Prasad Rao, and William Horne.
\newblock Detecting malicious domains via graph inference.
\newblock In {\em European Symposium on Research in Computer Security}, pages
  1--18. Springer, 2014.

\bibitem{fraudeagle}
Leman Akoglu, Rishi Chandy, and Christos Faloutsos.
\newblock Opinion fraud detection in online reviews by network effects.
\newblock {\em ICWSM}, 13:2--11, 2013.

\bibitem{generalbp}
Jonathan~S Yedidia, William~T Freeman, and Yair Weiss.
\newblock Understanding belief propagation and its generalizations.
\newblock {\em Exploring artificial intelligence in the new millennium},
  8:236--239, 2003.

\bibitem{enterprise_bp}
Alina Oprea, Zhou Li, Ting-Fang Yen, Sang~H Chin, and Sumayah Alrwais.
\newblock Detection of early-stage enterprise infection by mining large-scale
  log data.
\newblock In {\em 2015 45th Annual IEEE/IFIP International Conference on
  Dependable Systems and Networks}, pages 45--56. IEEE, 2015.

\bibitem{codaspy_issa}
Issa~M Khalil, Bei Guan, Mohamed Nabeel, and Ting Yu.
\newblock A domain is only as good as its buddies: Detecting stealthy malicious
  domains via graph inference.
\newblock In {\em Proceedings of the Eighth ACM Conference on Data and
  Application Security and Privacy}, pages 330--341. ACM, 2018.

\bibitem{fastflux:2008:NDSS}
Thorsten Holz, Christian Gorecki, Konrad Rieck, and Felix~C. Freiling.
\newblock {Measuring and Detecting Fast-Flux Service Networks}.
\newblock In {\em Proceedings of the 15th Network and Distributed System
  Security Symposium}, 2008.

\bibitem{takedown:2019:NDSS}
Eihal Alowaisheq, Peng Wang, Sumayah~A. Alrwais, Xiaojing Liao, XiaoFeng Wang,
  Tasneem Alowaisheq, Xianghang Mi, Siyuan Tang, and Baojun Liu.
\newblock Cracking the wall of confinement: Understanding and analyzing
  malicious domain take-downs.
\newblock In {\em 26th Annual Network and Distributed System Security
  Symposium, {NDSS}}, 2019.

\bibitem{impending}
Mahmood Sharif, Jumpei Urakawa, Nicolas Christin, Ayumu Kubota, and Akira
  Yamada.
\newblock Predicting impending exposure to malicious content from user
  behavior.
\newblock In {\em Proceedings of the 2018 ACM SIGSAC Conference on Computer and
  Communications Security}, pages 1487--1501. ACM, 2018.

\bibitem{robust}
Vincent~F Taylor, Riccardo Spolaor, Mauro Conti, and Ivan Martinovic.
\newblock Robust smartphone app identification via encrypted network traffic
  analysis.
\newblock {\em IEEE Transactions on Information Forensics and Security}, 2017.

\bibitem{guiltfile}
Acar Tamersoy, Kevin Roundy, and Duen~Horng Chau.
\newblock Guilt by association: large scale malware detection by mining
  file-relation graphs.
\newblock In {\em Proceedings of the 20th ACM SIGKDD international conference
  on Knowledge discovery and data mining}, pages 1524--1533. ACM, 2014.

\bibitem{lp:2002}
Xiaojin Zhu and Zoubin Ghahramani.
\newblock Learning from labeled and unlabeled data with label propagation.
\newblock 2002.

\bibitem{marmite}
Gianluca Stringhini, Yun Shen, Yufei Han, and Xiangliang Zhang.
\newblock Marmite: spreading malicious file reputation through download graphs.
\newblock In {\em Proceedings of the 33rd Annual Computer Security Applications
  Conference}, pages 91--102. ACM, 2017.

\bibitem{node2vec}
Aditya Grover and Jure Leskovec.
\newblock Node2vec: Scalable feature learning for networks.
\newblock In {\em Proceedings of the 22Nd ACM SIGKDD International Conference
  on Knowledge Discovery and Data Mining}, KDD '16, pages 855--864, New York,
  NY, USA, 2016. ACM.

\bibitem{appprint}
Stanislav Miskovic, Gene~Moo Lee, Yong Liao, and Mario Baldi.
\newblock Appprint: automatic fingerprinting of mobile applications in network
  traffic.
\newblock In {\em International Conference on Passive and Active Network
  Measurement}, pages 57--69. Springer, 2015.

\bibitem{tcp_finger}
Hasan~Faik Alan and Jasleen Kaur.
\newblock Can android applications be identified using only tcp/ip headers of
  their launch time traffic?
\newblock In {\em Proceedings of the 9th ACM conference on security \& privacy
  in wireless and mobile networks}, pages 61--66. ACM, 2016.

\bibitem{inconsistent}
Abner Mendoza and Guofei Gu.
\newblock Mobile application web api reconnaissance: Web-to-mobile
  inconsistencies \& vulnerabilities.
\newblock In {\em 2018 IEEE Symposium on Security and Privacy (SP)}, pages
  756--769. IEEE, 2018.

\bibitem{networkframe}
Arash~Habibi Lashkari, Andi Fitriah~A Kadir, Hugo Gonzalez, Kenneth~Fon Mbah,
  and Ali~A Ghorbani.
\newblock Towards a network-based framework for android malware detection and
  characterization.
\newblock In {\em Proceeding of the 15th international conference on privacy,
  security and trust}, 2017.

\bibitem{locationtracking}
Boyang Hu, Qicheng Lin, Yao Zheng, Qiben Yan, Matthew Troglia, and Qingyang
  Wang.
\newblock Characterizing location-based mobile tracking in mobile ad networks.
\newblock {\em arXiv preprint arXiv:1903.09916}, 2019.

\bibitem{ample}
Gyan Ranjan, Alok Tongaonkar, and Ruben Torres.
\newblock Approximate matching of persistent lexicon using search-engines for
  classifying mobile app traffic.
\newblock In {\em IEEE INFOCOM 2016-The 35th Annual IEEE International
  Conference on Computer Communications}, pages 1--9. IEEE, 2016.

\bibitem{lookat}
Alok Tongaonkar.
\newblock A look at the mobile app identification landscape.
\newblock {\em IEEE Internet Computing}, 20(4):9--15, 2016.

\bibitem{appcracker}
Fangda Cai, Hao Chen, Yuanyi Wu, and Yuan Zhang.
\newblock Appcracker: Widespread vulnerabilities in user and session
  authentication in mobile apps.
\newblock {\em MoST 2015}, 2015.

\bibitem{androidP}
Chad Brubaker.
\newblock Protecting users with tls by default in android p, 2018.

\bibitem{debate}
Elias~P. Papadopoulos, Michalis Diamantaris, Panagiotis Papadopoulos, Thanasis
  Petsas, Sotiris Ioannidis, and Evangelos~P. Markatos.
\newblock The long-standing privacy debate: Mobile websites vs mobile apps.
\newblock In {\em Proceedings of the 26th International Conference on World
  Wide Web}, pages 153--162, 2017.

\bibitem{mixcontent}
Google Developers.
\newblock Mixed content weakens https, 2019.

\bibitem{breaking}
Narseo Vallina-Rodriguez, Jay Shah, Alessandro Finamore, Yan Grunenberger,
  Konstantina Papagiannaki, Hamed Haddadi, and Jon Crowcroft.
\newblock Breaking for commercials: characterizing mobile advertising.
\newblock In {\em Proceedings of the 2012 Internet Measurement Conference},
  pages 343--356. ACM, 2012.

\bibitem{httpscost}
David Naylor, Alessandro Finamore, Ilias Leontiadis, Yan Grunenberger, Marco
  Mellia, Maurizio Munaf{\`o}, Konstantina Papagiannaki, and Peter Steenkiste.
\newblock The cost of the s in https.
\newblock In {\em Proceedings of the 10th ACM International on Conference on
  emerging Networking Experiments and Technologies}, pages 133--140. ACM, 2014.

\bibitem{nazca}
Luca Invernizzi, Stanislav Miskovic, Ruben Torres, Christopher Kruegel,
  Sabyasachi Saha, Giovanni Vigna, Sung-Ju Lee, and Marco Mellia.
\newblock Nazca: Detecting malware distribution in large-scale networks.
\newblock In {\em NDSS}, volume~14, pages 23--26. Citeseer, 2014.

\bibitem{dissecting}
Yajin Zhou and Xuxian Jiang.
\newblock Dissecting android malware: Characterization and evolution.
\newblock In {\em 2012 IEEE symposium on security and privacy}, pages 95--109.
  IEEE, 2012.

\bibitem{virustotal}
VirusTotal.
\newblock Virustotal, 2019.

\bibitem{unknown_malice}
Kai Chen, Peng Wang, Yeonjoon Lee, XiaoFeng Wang, Nan Zhang, Heqing Huang, Wei
  Zou, and Peng Liu.
\newblock Finding unknown malice in 10 seconds: Mass vetting for new threats at
  the google-play scale.
\newblock In {\em 24th $\{$USENIX$\}$ Security Symposium ($\{$USENIX$\}$
  Security 15)}, pages 659--674, 2015.

\bibitem{mastino_mining}
Babak Rahbarinia, Marco Balduzzi, and Roberto Perdisci.
\newblock Real-time detection of malware downloads via large-scale url file
  machine graph mining.
\newblock In {\em Proceedings of the 11th ACM on Asia Conference on Computer
  and Communications Security}, pages 783--794. ACM, 2016.

\bibitem{usagepattern}
Alok Tongaonkar, Shuaifu Dai, Antonio Nucci, and Dawn Song.
\newblock Understanding mobile app usage patterns using in-app advertisements.
\newblock In {\em International Conference on Passive and Active Network
  Measurement}, pages 63--72. Springer, 2013.

\bibitem{appriskanalysis}
Muhammad Ikram, Narseo Vallina-Rodriguez, Suranga Seneviratne, Mohamed~Ali
  Kaafar, and Vern Paxson.
\newblock An analysis of the privacy and security risks of android vpn
  permission-enabled apps.
\newblock In {\em Proceedings of the 2016 Internet Measurement Conference},
  pages 349--364. ACM, 2016.

\bibitem{droppereffect}
Bum~Jun Kwon, Jayanta Mondal, Jiyong Jang, Leyla Bilge, and Tudor
  Dumitra{\c{s}}.
\newblock The dropper effect: Insights into malware distribution with
  downloader graph analytics.
\newblock In {\em Proceedings of the 22nd ACM SIGSAC Conference on Computer and
  Communications Security}, pages 1118--1129. ACM, 2015.

\bibitem{polonium}
Duen~Horng Chau, Carey Nachenberg, Jeffrey Wilhelm, Adam Wright, and Christos
  Faloutsos.
\newblock Polonium: Tera-scale graph mining and inference for malware
  detection.
\newblock In {\em Proceedings of the 2011 SIAM International Conference on Data
  Mining}, pages 131--142. SIAM, 2011.

\bibitem{speagle}
Shebuti Rayana and Leman Akoglu.
\newblock Collective opinion spam detection: Bridging review networks and
  metadata.
\newblock In {\em Proceedings of the 21th acm sigkdd international conference
  on knowledge discovery and data mining}, pages 985--994. ACM, 2015.

\bibitem{getoff}
Yajin Zhou, Zhi Wang, Wu~Zhou, and Xuxian Jiang.
\newblock Hey, you, get off of my market: detecting malicious apps in official
  and alternative android markets.
\newblock In {\em NDSS}, volume~25, pages 50--52, 2012.

\bibitem{snare}
Mary McGlohon, Stephen Bay, Markus~G Anderle, David~M Steier, and Christos
  Faloutsos.
\newblock Snare: a link analytic system for graph labeling and risk detection.
\newblock In {\em Proceedings of the 15th ACM SIGKDD international conference
  on Knowledge discovery and data mining}, pages 1265--1274. ACM, 2009.

\bibitem{exposure}
Leyla Bilge, Engin Kirda, Christopher Kruegel, and Marco Balduzzi.
\newblock Exposure: Finding malicious domains using passive dns analysis.
\newblock In {\em Ndss}, pages 1--17, 2011.

\bibitem{edgepotential}
Jaemin Yoo, Saehan Jo, and U~Kang.
\newblock Supervised belief propagation: Scalable supervised inference on
  attributed networks.
\newblock In {\em Data Mining (ICDM), 2017 IEEE International Conference on},
  pages 595--604. IEEE, 2017.

\bibitem{alexa}
Alexa.
\newblock Alexa top sites, 2019.

\bibitem{trafficgraph}
Jinjun Tang, Yinhai Wang, Hua Wang, Shen Zhang, and Fang Liu.
\newblock Dynamic analysis of traffic time series at different temporal scales:
  A complex networks approach.
\newblock {\em Physica A: Statistical Mechanics and its Applications},
  405:303--315, 2014.

\bibitem{droidjust}
Xin Chen and Sencun Zhu.
\newblock Droidjust: Automated functionality-aware privacy leakage analysis for
  android applications.
\newblock In {\em Proceedings of the 8th ACM Conference on Security \& Privacy
  in Wireless and Mobile Networks}, page~5. ACM, 2015.

\bibitem{credroid}
Jyoti Malik and Rishabh Kaushal.
\newblock Credroid: Android malware detection by network traffic analysis.
\newblock In {\em Proceedings of the 1st ACM Workshop on Privacy-Aware Mobile
  Computing}, pages 28--36. ACM, 2016.

\bibitem{reg:2017:RAID}
Thomas Vissers, Jan Spooren, Pieter Agten, Dirk Jumpertz, Peter Janssen, Marc
  Van~Wesemael, Frank Piessens, Wouter Joosen, and Lieven Desmet.
\newblock Exploring the ecosystem of malicious domain registrations in the .eu
  tld.
\newblock In {\em Research in Attacks, Intrusions, and Defenses}, pages
  472--493. Springer International Publishing, 2017.

\bibitem{DNSDB}
{Farsight Security, Inc.}
\newblock {DNS Database}.
\newblock \url{https://www.dnsdb.info/}, 2019.
\newblock Accessed: 15-05-2019.

\bibitem{realtime}
Junming Liu, Yanjie Fu, Jingci Ming, Yong Ren, Leilei Sun, and Hui Xiong.
\newblock Effective and real-time in-app activity analysis in encrypted
  internet traffic streams.
\newblock In {\em Proceedings of the 23rd ACM SIGKDD International Conference
  on Knowledge Discovery and Data Mining}, KDD '17, pages 335--344, New York,
  NY, USA, 2017. ACM.

\bibitem{androzoo}
Kevin Allix, Tegawend{\'e}~F. Bissyand{\'e}, Jacques Klein, and Yves Le~Traon.
\newblock Androzoo: Collecting millions of android apps for the research
  community.
\newblock In {\em Proceedings of the 13th International Conference on Mining
  Software Repositories}, MSR '16, pages 468--471, New York, NY, USA, 2016.
  ACM.

\bibitem{koodous}
Koodous.
\newblock Koodous: Online malware analysis platform, 2019.

\bibitem{droidminer}
Chao Yang, Zhaoyan Xu, Guofei Gu, Vinod Yegneswaran, and Phillip Porras.
\newblock Droidminer: Automated mining and characterization of fine-grained
  malicious behaviors in android applications.
\newblock In {\em European symposium on research in computer security}, pages
  163--182. Springer, 2014.

\bibitem{drebin}
Daniel Arp, Michael Spreitzenbarth, Malte Hubner, Hugo Gascon, Konrad Rieck,
  and CERT Siemens.
\newblock Drebin: Effective and explainable detection of android malware in
  your pocket.
\newblock In {\em Ndss}, volume~14, pages 23--26, 2014.

\bibitem{droidscope}
Lok-Kwong Yan and Heng Yin.
\newblock Droidscope: Seamlessly reconstructing the os and dalvik semantic
  views for dynamic android malware analysis.
\newblock In {\em USENIX security symposium}, pages 569--584, 2012.

\bibitem{automated}
Apostolis Zarras, Antonis Papadogiannakis, Robert Gawlik, and Thorsten Holz.
\newblock Automated generation of models for fast and precise detection of
  http-based malware.
\newblock In {\em Privacy, Security and Trust (PST), 2014 Twelfth Annual
  International Conference on}, pages 249--256. IEEE, 2014.

\bibitem{androidmalware}
Anshul Arora, Shree Garg, and Sateesh~K Peddoju.
\newblock Malware detection using network traffic analysis in android based
  mobile devices.
\newblock In {\em Next generation mobile apps, services and technologies
  (NGMAST), 2014 eighth international conference on}, pages 66--71. IEEE, 2014.

\bibitem{Yuan:2019:fakeaccounts}
Dong Yuan, Yuanli Miao, Neil~Zhenqiang Gong, Zheng Yang, Qi~Li, Dawn Song, Qian
  Wang, and Xiao Liang.
\newblock Detecting fake accounts in online social networks at the time of
  registrations.
\newblock In {\em Proceedings of the 2019 ACM SIGSAC Conference on Computer and
  Communications Security}, CCS '19, pages 1423--1438, New York, NY, USA, 2019.
  ACM.

\end{thebibliography}

\section*{Appendix}
\begin{table}[h]%%
\begin{tabular}{|p{1.5cm}|p{3cm}|p{10cm}|}\hline
\centering
\textbf{Category}  & \textbf{Type} & \textbf{Example Keywords}\\
\hline
 & Sim card number & iccid, simserialnumber, simno, simnumber, sim\\ \cline{2-3}
Phone & IMSI & imsi, mobileimsi, user-imsi, imsi1, imsi\_no, x-imsi, client-imsi\\\cline{2-3}
 & Phone number & phone\_num, phone, tel\_num, mobile\_no, cellphonenumber, cellphone, userphone, tel\_number, usrphonenum\\
\hline
& User ID & user\_nick, user\_id, user\_name, userid, x-userid, log-user-id, login\_name, client-user-id\\ \cline{2-3}
User        & Email & user\_email, email, login\_email, email\_name, contact\_email, acc\_email\\\cline{2-3}
& Birthday & birthday, user\_birth, customer\_birthday, passenger\_birthday\\ \cline{2-3}
& Gender & gender, sex, client\_gender\\
\hline
Credential        & Password & userpwd, password, passwd, user\_pwd, \_password, pwd\\
\hline
Location    & Location & longitude, latitude, lng\_lat, coordinate, homeaddress, coords, geo\_location, gps\_long, geoInfo\\
\hline
 & UDID & udid, device\_udid\\ \cline{2-3}
 & Device ID & devid, device\_id, x-device-id, deviceid\\\cline{2-3}
 Device & IMEI & imei, device-imei, phone-imei, imei1, mobileimei\\\cline{2-3}
 Identifier & GUID & guid, phoneguid, dev-guid\\\cline{2-3}
 & UUID & uuid, device\_uuid, phoneuuid, x-device-uuid, uuid2\\\cline{2-3}
 & MAC & user\_mac, mac, mac\_addr, \_mac, x-macaddress\\\cline{2-3}
 & Android ID & android\_id, androidid, \_androidid, androidid1\\
\hline
\end{tabular}
\vspace{.12in}
\caption{List of private information, type, and example keywords}
\label{pii_keywords}

\end{table}

\end{document}